\documentclass[aps,pra,amsmath,showpacs,preprintnumbers,twocolumn,floatfix,amssymb]{revtex4}
\usepackage{graphicx,color}
\usepackage{braket}
\usepackage{color}
\DeclareGraphicsExtensions{.eps,.ps}

\usepackage{epstopdf}
\usepackage[latin1]{inputenc}
\usepackage{amsmath,amsbsy,amsfonts,amssymb}
\usepackage{bm}
\usepackage{epsfig}
\usepackage{dcolumn}

\newcommand{\Xstate}{\mbox{$X^1\Sigma^+$}}
\newcommand{\astate}{\mbox{$a^3\Sigma^+$}}
\newcommand{\wn}{cm$^{-1}$}

\begin{document}

\title{Towards the production of ultracold ground-state RbCs molecules: \\
Feshbach resonances, weakly bound states, and coupled-channel model}

\author{Tetsu Takekoshi$^{1,2}$}
\author{Markus Debatin$^{1}$}
\author{Raffael Rameshan$^{1}$}
\author{Francesca Ferlaino$^{1}$}
\author{Rudolf Grimm$^{1,2}$}
\author{Hanns-Christoph N\"{a}gerl$^{1}$}
\affiliation{
$^{1}$Institut f\"ur Experimentalphysik, Universit\"at Innsbruck, 6020 Innsbruck, Austria
\\
$^{2}$Institut f\"ur Quantenoptik und Quanteninformation,
\"Osterreichische Akademie der Wissenschaften, 6020 Innsbruck, Austria
}
\author{C.~Ruth Le Sueur}
\author{Jeremy M. Hutson}
\affiliation{Department of Chemistry, Durham University, South Road,
Durham DH1 3LE, United Kingdom}
\author{Paul S. Julienne}
\affiliation{Joint Quantum Institute, NIST and University of Maryland,
Gaithersburg, MD 20899, USA}
\author{Svetlana Kotochigova}
\affiliation{Physics Department, Temple University, Philadelphia, PA 19122-6082, USA}
\author{Eberhard Tiemann}
\affiliation{Institute of Quantum Optics, Leibniz Universit\"{a}t
Hannover, D-30167 Hannover, Germany}

\date{\today}

\begin{abstract}
We have studied interspecies scattering in an ultracold mixture of
$^{87}$Rb and $^{133}$Cs atoms, both in their lowest-energy spin
states. The three-body loss signatures of 30 incoming $s$- and $p$-wave
magnetic Feshbach resonances over the range 0 to 667 G have been
catalogued. Magnetic field modulation spectroscopy was used to observe
molecular states bound by up to 2.5 MHz$\times h$. We have created RbCs
Feshbach molecules using two of the resonances. Magnetic moment
spectroscopy along the magneto-association pathway from 197 to 182~G
gives results consistent with the observed and calculated dependence of
the binding energy on magnetic field strength. We have set up a
coupled-channel model of the interaction and have used direct
least-squares fitting to refine its parameters to fit the experimental
results from the Feshbach molecules, in addition to the Feshbach
resonance positions and the spectroscopic results for deeply bound
levels. The final model gives a good description of all the
experimental results and predicts a large resonance near 790 G, which
may be useful for tuning the interspecies scattering properties.
Quantum numbers and vibrational wavefunctions from the model can also
be used to choose optimal initial states of Feshbach molecules for
their transfer to the rovibronic ground state using stimulated Raman
adiabatic passage (STIRAP).
\end{abstract}

%     cooling mol.;   trap molec.;   atom in lattice      Quant opt.
% PACS, the Physics and Astronomy Classification Scheme.
\maketitle

\section{Introduction}
\label{intro}

Dilute quantum gases are ideal for studying many-body physics, because
they provide model systems in which the parameters can be precisely
controlled. External fields can be used to tune the effective isotropic
contact interactions between the particles, and the geometry and
strength of the confining optical potentials can be controlled by laser
beams. For example, quantum-gas analogues of superconductivity
\cite{Zwierlein:2005} and the superfluid-to-Mott-insulator quantum
phase transition \cite{Greiner:2002} have been observed in the
laboratory, and their properties have been shown to agree beautifully
with the predictions from theoretical models \cite{Giorgini:2008}.
Recently, quantum gases of particles with long-range anisotropic
interactions have been created \cite{Griesmaier:2005, Griesmaier:2006,
Ni:KRb:2008}. For particles with permanent electric dipole moments, the
range of the dipole-dipole interactions can be much larger than typical
optical lattice spacings, and interesting new quantum phases and
quantum information applications have been proposed \cite{Goral:2002,
Buechler:2007, Micheli:2007, Pupillo:2008, Wall:2009}. A quantum gas of
$^{40}$K$^{87}$Rb ground-state molecules is the only such system that
presently exists in the laboratory \cite{Ni:KRb:2008}.

Our goal is to generate a dipolar quantum gas of ground-state
$^{87}$Rb$^{133}$Cs, which, unlike KRb, is expected to be collisionally
stable because both the exchange reaction 2RbCs $\rightarrow$ Rb$_2$ +
Cs$_2$ and trimer formation reactions are endothermic
\cite{Zuchowski:trimers:2010}. Although other approaches are under
development \cite{Bethlem:IRPC:2003, Sage:2005, Deiglmayr:2008,
Aikawa2010cto}, the only method currently available to produce high
phase-space density gases of ground-state molecules is to create weakly
bound molecules from ultracold atomic gases by magnetic tuning across a
Feshbach resonance \cite{Regal2003cum, Herbig2003poa}, and then to
transfer the molecules to the rovibronic ground state by stimulated
Raman adiabatic passage (STIRAP) \cite{Bergmann1998cpt, Winkler2007cot,
Danzl2008qgo, Ni:KRb:2008, Lang2008utm, Mark2009drf,
Danzl:ground:2010}. As a first step, we have performed evaporative
cooling on Rb and Cs samples in separate optical traps, combining them
at the end to obtain an Rb-Cs mixture with high phase-space density
\cite{Lercher:2011}. We have successfully used this mixture to produce
ultracold samples of weakly bound RbCs \cite{Debatin:2011}. In this
paper, we present a combined experimental and theoretical study of the
interspecies Feshbach resonances and weakly bound molecular energy
levels of Rb-Cs and use the results to develop an accurate
coupled-channel model of the interaction, based on the derived
interaction potentials for the molecular states $X\,^1\Sigma^+$ and $a
\, ^3\Sigma^+$.

\section{Overview}

The work described in this paper involved a close collaboration between
experiment and theory. At the start of the work, the Feshbach
resonances and bound states observed experimentally \cite{Pilch:2009}
were unassigned. In initial theoretical work, we developed preliminary
coupled-channel models of the bound states and scattering and used
these to propose assignments of quantum numbers to observed energy
levels and Feshbach resonances. Experiments were then carried out to
test the assignments and extend the early measurements. The whole
process was repeated several times. However, to aid understanding, we
will describe the experiments in Section III below using quantum
numbers based on our final understanding from theory (Section IV), even
though the quantum numbers were not known at the outset.

%%%%%%%%%%%%%%%%%%%%%%%%%%%%%%%%%
\begin{figure}[tbp]
\includegraphics[angle=-90,width=\columnwidth]{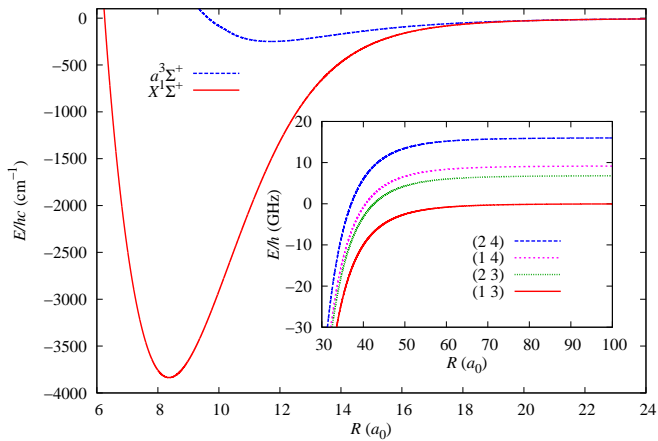}

\caption{[Color online.] Molecular potential energy curves $V_0(R)$ and
$V_1(R)$ for the singlet and triplet states of RbCs correlating with
two separated $^2S_{1/2}$ ground-state atoms. The inset shows an
expanded view of the long-range potentials separating to the four
different hyperfine thresholds at zero field, labelled by
$(f_{^{87}{\rm Rb}},f_{^{133}{\rm Cs}})$, but no longer by singlet or
triplet.} \label{fig-curves}
\end{figure}
%%%%%%%%%%%%%%%%%%%%%%%%%%%%%%%%%

Two alkali-metal atoms in $^2S$ states interact at short range to form
singlet ($X\,^1\Sigma^+$) and triplet ($a\,^3\Sigma^+$) states. Docenko
{\em et al.}\ \cite{Docenko:RbCs:2011} have carried out an extensive
spectroscopic study of these states by Fourier transform spectroscopy
and have developed potential energy curves as shown in Fig.\
\ref{fig-curves}. They were able to observe the vibrational ladder up
to high-lying levels with outer turning points around 1.5 nm, at which
point the coupling between singlet and triplet molecular states is
already significant. They also identified in the observed spectra
accidental coincidences of singlet and triplet levels deeper within the
potential wells, which fixed their relative energy position very well.
In the present work, we initially constrained the short-range part of
the potential to follow these curves and adjusted the long-range
parameters to reproduce the Feshbach resonances and weakly bound
states.

The bound states (Feshbach molecules) that are of most interest in the
present paper have binding energies of at most a few MHz$\times h$
\footnote{We use units of energy and frequency interchangeably in the
text, in accordance with the conventional usage in this field of
physics.} and require a quite different description. For a
heteronuclear bialkali molecule, there are 4 field-free atomic
thresholds, which for $^{87}$Rb$^{133}$Cs may be labelled in increasing
order of energy by $(f_{\rm Rb},f_{\rm Cs})$ = (1,3), (2,3), (1,4), and
(2,4), as shown in the inset of Fig.~\ref{fig-curves}. In a magnetic
field, each threshold splits into $(2f_{\rm Rb}+1)(2f_{\rm Cs}+1)$
sublevels labelled $|f_{\rm Rb},m_{\rm Rb}\rangle + |f_{\rm Cs},m_{\rm
Cs}\rangle$. The Feshbach molecules might be described using two
different sets of quantum numbers, either $(f_{\rm Rb},m_{\rm
Rb},f_{\rm Cs},m_{\rm Cs})$ or $(f_{\rm Rb},f_{\rm Cs},F,M_F)$, where
$F$ is the resultant of $f_{\rm Rb}$ and $f_{\rm Cs}$ and $M_F=m_{\rm
Rb}+m_{\rm Cs}$. In the non-rotating case $F$ and $M_F$ are exact
quantum numbers if there is no external field, but if there is an
external magnetic field, it mixes states with different $F$ values,
destroying the exactness of $F$ as a quantum number; the character of
the Feshbach molecules at the magnetic fields considered here is more
accurately described by $(f_{\rm Rb},m_{\rm Rb},f_{\rm Cs},m_{\rm
Cs})$.  For high magnetic fields, $f_{\rm Rb}$ and $f_{\rm Cs}$ are
also no longer good quantum numbers.

%%%%%%%%%%%%%%%%%%%%%%%%%%%%%%%%%
\begin{figure}[tbp]

\includegraphics[width=\columnwidth]{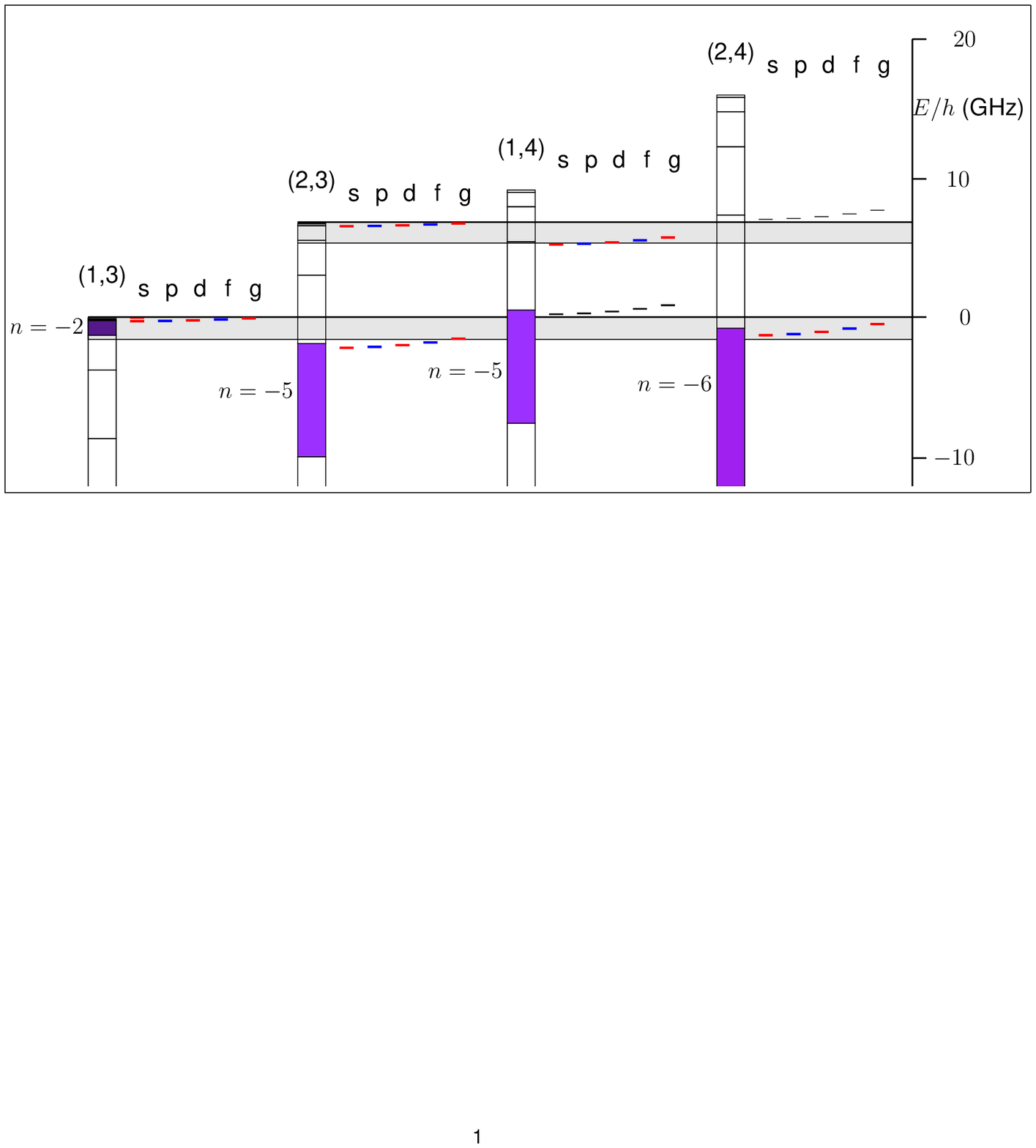}
\caption{[Color online.] Thresholds for $^{87}$Rb$^{133}$Cs and the
``bins" (vertical boxes) below each threshold within which each
vibrational state must lie for any value of the scattering length. The
horizontal boxes, 1.5 GHz deep, show the energy range within which
bound levels can cause resonances at the $|1,1\rangle + |3,3\rangle$
and $|2,-1\rangle + |3,3\rangle$ thresholds at fields under 500~G. The
vibrational ``bin'' for each hyperfine state that contains bound states
that can cause resonances at the lowest threshold is colored.  Selected
levels are shown (short horizontal lines) as a function of $L$ ($s$,
$p$, $d$, $f$, $g$ for $L=0$, 1, 2, 3, 4) for the specific choice of
scattering length that gives a least-bound state for $L=0$ with 110(20)
kHz below threshold. For fields below 500 G, only levels $|n(f_{\rm
Rb},f_{\rm Cs})\rangle=|{-2}(1,3)\rangle$ and $|{-6}(2,4)\rangle$ can
cause resonances at the $|1,1\rangle + |3,3\rangle$ threshold and only
levels $|{-4}(1,4)\rangle$ and $|{-2}(2,3)\rangle$ can cause resonances
at the $|2,-1\rangle + |3,3\rangle$ threshold.}
\label{thresholds_and_bins}
\end{figure}
%%%%%%%%%%%%%%%%%%%%%%%%%%%%%%%%%

Additional quantum numbers are needed for the molecules' end-over-end angular
momentum $L$ and the molecular vibration. For near-dissociation levels
it is convenient to specify the vibrational quantum number with respect
to the asymptote of the atom pair, so that the topmost level is $n=-1$,
the next is $n=-2$, and so on. Each level lies within a ``bin" below
its associated dissociation threshold, with the boundaries of the bins
determined by the long-range forces between the atoms. For RbCs, using
the published values of the long-range dispersion coefficients,
\cite{Derevianko:2001,Porsev:2003}, which are the same for the
singlet and triplet potentials, we find that the $n=-1$ level lies
between zero and $-165$ MHz, and the $n=-2$ level lies between $-165$
and $-1150$ MHz. Subsequent lower bin boundaries lie at 3.7, 8.6, 16.7
and 28.8 GHz below threshold for $n=-3$ to $-6$, respectively. As shown
below, the actual levels for $^{87}$RbCs lie close to the top of their
bins.

Feshbach resonances occur at fields where a bound state exists at the
same energy as the colliding atoms. Zero-energy Feshbach resonances are
caused by molecular levels that cross atomic thresholds as a function
of magnetic field. Since the level shifts due to the Zeeman effect at
fields below 500~G \footnote{Units of Gauss rather than Tesla, the
accepted SI unit for the magnetic field, have been used in this paper to
conform to the conventional usage in this field of physics.} are not
more than 1.5 GHz, there is only one vibrational level below each
field-free threshold that can cause Feshbach resonances at the
$|1,1\rangle + |3,3\rangle$ threshold, as shown in Fig.\
\ref{thresholds_and_bins}; these are $n=-5$, $-5$ and $-6$ for levels
associated with $(f_{\rm Rb},f_{\rm Cs}) = (2,3)$, (1,4) and (2,4),
respectively. In addition to this, levels very close to dissociation
($n=-1$ or $-2$) corresponding to the {\em same} zero-field threshold
as the incoming wave can also cause low-field resonances. Fig.\
\ref{thresholds_and_bins} also shows the situation at the $|2,-1\rangle
+ |3,3\rangle$ threshold, which will be considered in Section
\ref{sec:excited}.

In the general case we label weakly bound states with a complete set of
quantum numbers $|n(f_{\rm Rb},f_{\rm Cs})L(m_{\rm Rb},m_{\rm Cs}),
M\rangle$, with $L=0, 1, 2$, etc.\ designated by $s$, $p$, $d$, etc.
$M$ is the sum of all angular momenta projected onto the field axis,
$M=m_{\rm Rb}+m_{\rm Cs}+M_L$, and is the only exactly conserved
quantum number in an external field.  Since, however, $M$ is always 4
for the levels studied in this paper (except in sections
\ref{sec:p-wave} and \ref{sec:excited}), we will omit it in the
following discussion. All other angular momenta are approximate quantum
numbers, but are sufficient for proper labeling. We characterize by $L_\text{c}=0,1,...$ the partial wave character of the continuum scattering process and speak of incoming $s$- and $p$-wave resonances for $L_\text{c}=0$ and $L_\text{c}=1$, respectively.

\section{The Experiments}
\subsection{Feshbach resonances}\label{FR}
Magnetic Feshbach resonances are an important tool for the production
of weakly bound molecules and for tuning the scattering length, which
determines the elastic and inelastic scattering properties of cold
atomic gases \cite{Chin:RMP:2010}. In addition, their positions provide
important clues to the molecular bound state structure that lies below
the scattering threshold. In previous work \cite{Pilch:2009} we
observed 23 resonances over the range 0 to 300 G, using a mixture of
the lowest spin states, $^{87}$Rb$|1,1\rangle$ and
$^{133}$Cs$|3,3\rangle$. Since this mixture was prepared by evaporating
both species simultaneously in the same optical trap, interspecies
three-body recombination loss and heating \cite{Chin:RMP:2010} limited
the evaporative cooling efficiency, resulting in comparatively high
temperatures of 7~$\mu$K and low particle densities of about
$5\times10^{11}$ cm$^{-3}$ for each species.

\begin{figure}[htbp]
\includegraphics[width=8.6cm] {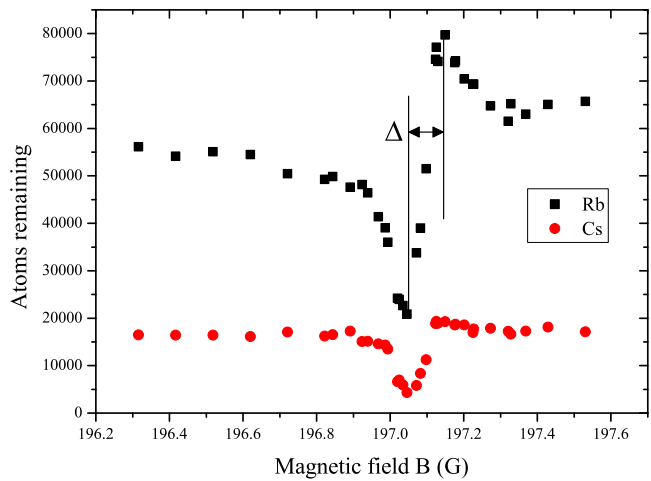}
\caption{[Color online.] Example of a Feshbach resonance scan showing
simultaneous loss for Rb and Cs near 197.06(5) G for a hold time of
$200$ ms. The maximum for the number of Rb atoms to the right of the
resonance is attributed to the zero crossing for the interspecies
scattering length. The difference in the positions of the minima and
the maximum as indicated gives an estimate for the resonance width
$\Delta$.}
\label{Feshbachraw}
\end{figure}

In the current experiment, the mixture is created by combining
separately cooled atomic clouds \cite{Lercher:2011}, so it is much
colder (100 to 200 nK) and denser ($5\times10^{12}$ cm$^{-3}$) and
gives a much better signal-to-noise ratio for the loss features
discussed below. We stop the evaporation procedure before the onset of
condensation, because we have previously found the two Bose-Einstein
condensates (BECs) to be immiscible \cite{Lercher:2011}. We hold the
mixture at constant magnetic field $B$ for 200 ms. Enhanced losses that
occur simultaneously for Rb and Cs are attributed to three-body
recombination \cite{Chin:RMP:2010} at an interspecies Feshbach
resonance. We associate the field value $B$ at which maximum atom loss
occurs with the pole of the resonance. For example,
Fig.~\ref{Feshbachraw} shows the atom loss in the vicinity of the
resonance near $197$ G.

For sufficiently wide resonances, we find that the number of Rb atoms
exhibits a maximum at fields just above resonance. Rb has a lower trap
depth than Cs and thus bears most of the heat load through evaporation
when the two species are in thermal equilibrium \cite{Lercher:2011}.
Reduced thermalization with Cs at zero interspecies scattering length
reduces the heat load on the Rb part of the sample and thus leads to
less loss of Rb atoms. This simple explanation allows us to provide an
estimate for the resonance width $\Delta$: It is the difference between
the field values for the minima (for Rb and Cs) and the maximum (for
Rb) as indicated in Fig.~\ref{Feshbachraw}. A detailed comparison with
calculated widths requires a thorough analysis, including three-body
and evaporation effects, and will be made in a future publication.

%%%%%%%%%%%%%%%%%%%%%%%%%%%%%%%%%
%%%%%%%%%%%%%%%%%%%%%%%%%%%%%%%%%

As part of this work, we have scanned over a wider range (0 to 667~G)
than in Ref.~\cite{Pilch:2009}, finding 7 new incoming $s$-wave
resonances in addition to those reported in Ref.~\cite{Pilch:2009}. The
old and new resonances are collected together in Table
\ref{resonances}. The resonances observed for temperatures $\leq 200$
nK are assigned as incoming $s$-wave ($L_\text{c} \!=\!0$) resonances, while
those observed at 7~$\mu$K and not observed at 200 nK are assigned as
incoming $p$-wave ($L_\text{c} \!=\!1$) resonances. The magnetic field is
calibrated near each resonance using Rb microwave transitions. The
calibration in our previous work \cite{Pilch:2009} was based on
low-field data and was found to deviate from the current calibration by
as much as 0.5 G when extrapolated to 300 G. The positions of the
incoming $p$-wave resonances have therefore been scaled to the new
calibration, using the incoming $s$-wave resonances observed in both
experiments as a reference. The incoming $p$-wave resonances from 258
to 272 G have been remeasured at $4\ \mu$K with the new calibration.

\begin{table}[htpb]
\caption{$^{87}$Rb$|1,1\rangle$ + Cs$|3,3\rangle$ Feshbach resonances
observed over the range 0--667~G for $s$-wave scattering and 0--300~G
for $p$-wave scattering. The magnetic field uncertainties result from a
quadrature of resonance position uncertainty due to atom number noise,
and an estimated field calibration error of 0.03~G. Resonances too
narrow to allow a clear width measurement have no width indicated.
\label{resonances}}
\begin{ruledtabular}
\begin{tabular}{ll||l}
\multicolumn{2}{c||}{$s$-wave}&$p$-wave\\
Field $B$~(G)&Width $\Delta$~(G)\qquad\qquad\qquad&Field $B$~(G)\\
% Lines of table here ending with \\
\hline
181.64(8)&0.27(10)&128.00(25)$^*$\\
197.06(5)&0.09(1)&129.60(25)$^*$\\
217.34(5)&0.06(1)&140.00(25)$^*$\\
225.43(3)&0.16(1)&140.50(25)$^*$\\
242.29(5)&&234.35(25)$^*$\\
247.32(5)&0.09(3)&235.96(25)$^*$\\
272.80(4)&&258.10(11)\\
273.45(4)&&259.60(11)\\
273.76(4)&&264.19(11)\\
279.12(5)&0.09(3)&266.23(11)\\
286.76(5)&&271.73(11)\\
308.44(5)&&289.97(25)$^*$\\
310.69(6)&0.60(4)&292.08(25)$^*$\\
314.74(11)&0.18(10)\\
352.65(34)&2.70(47)\\
381.34(5)&\\
421.93(5)&\\
\end{tabular}
\end{ruledtabular}
$^*$From Ref.~\cite{Pilch:2009} with field rescaled to current
calibration.\linebreak
\end{table}

%%%%%%%%%%%%%%%%%%%%%%%%%%%%%%%%%
\begin{figure}[htbp]
\includegraphics[width=8.6cm] {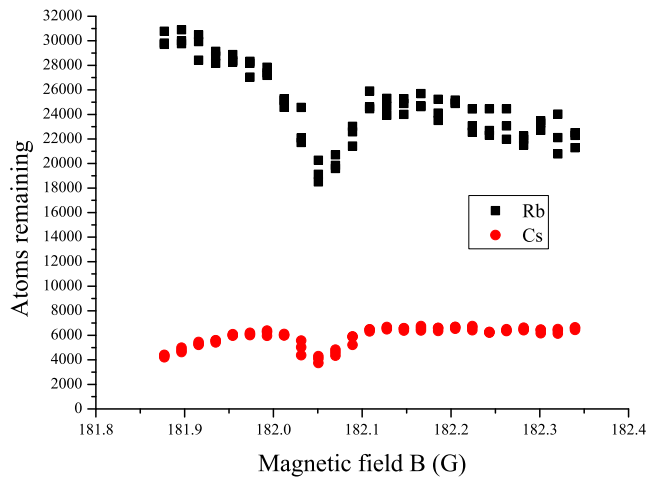}
\caption{[Color online.] Measurement of the binding energy of the Feshbach molecules. This is an example of a field modulation free-bound
resonance scan showing simultaneous loss for Rb and Cs for a hold time
of $900$ ms. The modulation frequency is held fixed at
$f_\text{m}=330$~kHz.} \label{wiggleraw}
\end{figure}
%%%%%%%%%%%%%%%%%%%%%%%%%%%%%%%%%

%%%%%%%%%%%%%%%%%%%%%%%%%%%%%%%%%
\begin{figure*}[tbp]
\includegraphics[angle=-90,width=2\columnwidth]{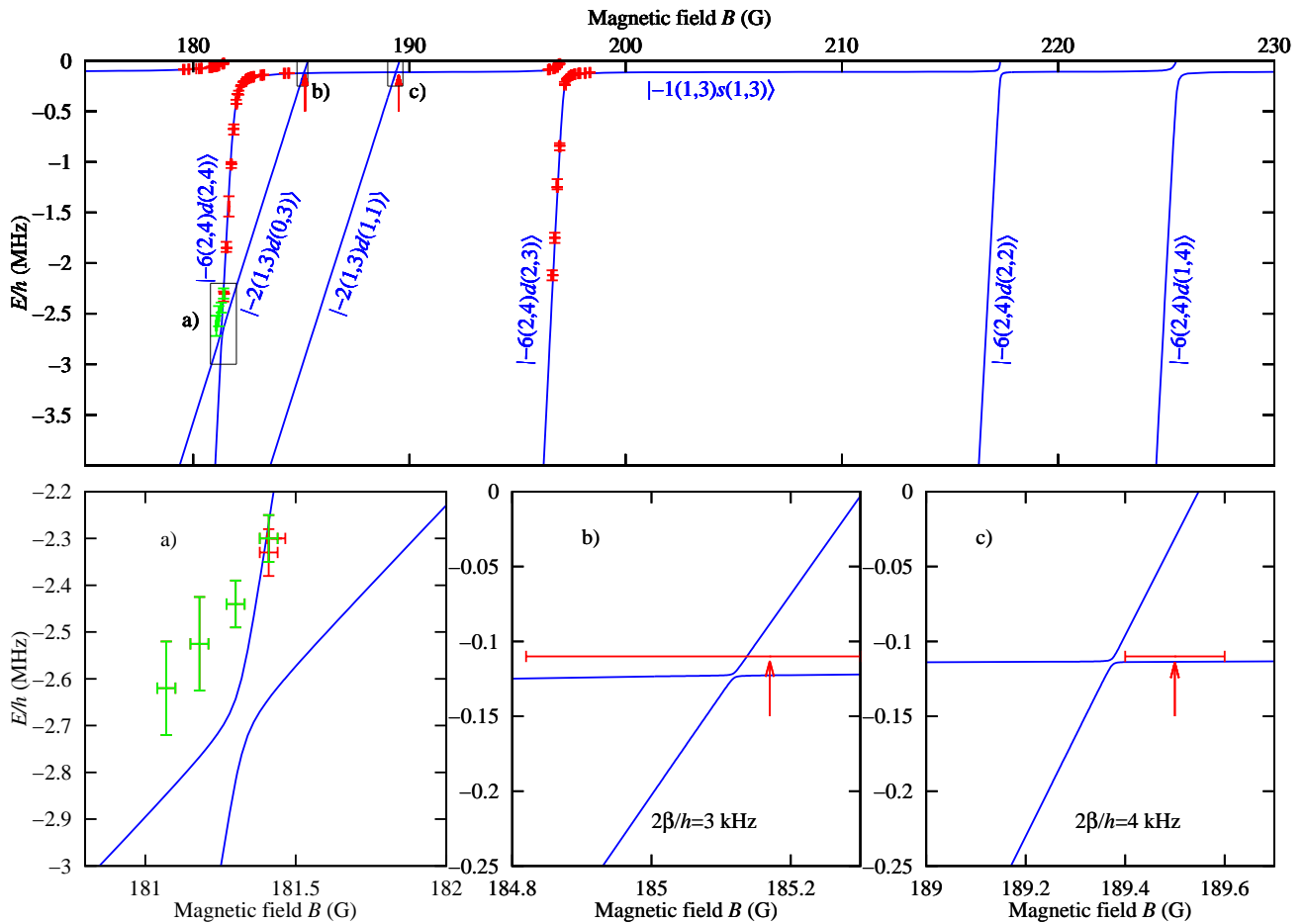}
\caption{[Color online.] Weakly bound states of RbCs obtained by
free-bound (red) and bound-free (green) magnetic-field modulation
spectroscopy, together with levels calculated for the final, fitted
potentials (blue solid lines). All levels are shown relative
to dissociation limit Rb $|1,1\rangle$ + Cs $|3,3\rangle$ at the given
magnetic field value $B$. Avoided crossings between the least-bound
state and the ramping $n=-2$ levels are shown as red arrows. The
smaller panels below, labeled a), b), and c) refer to the areas on the
larger panel marked by rectangles and the same labels. The quantity
$2\beta$ is the minimum separation in energy between the two states.}
\label{wiggle}
\end{figure*}
%%%%%%%%%%%%%%%%%%%%%%%%%%%%%%%%%

\subsection{Magnetic-field modulation spectroscopy}\label{MS}

We have used magnetic-field modulation spectroscopy \cite{Regal2003cum,
Thompson2005ump, Weber2008aou, Lange:2009} on our atom mixture to
measure binding energies of Feshbach molecules. A set of auxiliary
coils modulates the magnetic field $B$ along the quantization axis by
up to 0.2~G. Atom losses occur when the modulation frequency
$f_\text{m}$ is resonant with a free-bound transition
(Fig.~\ref{wiggleraw}). We observe the losses by holding $f_\text{m}$
fixed and scanning $B$, or by holding $B$ fixed and scanning
$f_\text{m}$. We find that the free-bound signal dies off for
$f_\text{m}$ above 2.5~MHz and attribute this to lower field amplitudes
generated by the coils due to their increased impedance at high
frequencies and to the Bessel-function-squared dependence of the
coupling strength on the binding energy \cite{Beaufils2010rfa}.

The binding energies obtained in this way near the Feshbach resonances
at 181.6~G and 197~G are plotted in Fig.~\ref{wiggle}. Two avoided
crossings close below threshold can clearly be identified. We attribute
these to the presence of a bound state running parallel to the atomic
threshold (with the same magnetic moment as the atom pair) with a
binding energy of approximately $110\pm20$~kHz$\times h$. This
``least-bound state" $|n(f_{\rm Rb}f_{\rm Cs})L(m_{\rm Rb}m_{\rm
Cs})\rangle=|{-1}(1,3)s(1,3)\rangle$ cannot be observed directly with
the modulation technique except near avoided crossings, because the
initial and final states involved are exactly equal in all spin quantum
numbers; they thus do not differ in magnetic moment and magnetic dipole
transitions between them are forbidden. The least-bound state causes
avoided crossings directly below the Feshbach resonances by the same
coupling mechanism as the Feshbach resonances, and the resulting mixed
states can be observed near these crossings. The binding energy of the
least-bound state allows us to estimate the interspecies background
scattering length as $+645(60)\ a_0$ for this scattering channel. This
value is further refined in Section \ref{sec:final} below. The large
value for the scattering length is responsible for the large background
interspecies thermalization and three-body loss rates observed
previously \cite{Lercher:2011,Cho2011,McCarron2011}.

\subsection{Feshbach molecules}\label{FM}
To create Feshbach molecules, we sweep the magnetic field $B$
adiabatically from high to low field across one of the Feshbach
resonances. The weakly bound molecules formed in this way can collide
with atoms and decay to deeply bound states. We must therefore remove
the atoms quickly. In previous experiments it was found that the atoms
can be removed from the molecular cloud with radiation pressure from a
laser (see e.g.~Refs.\ \cite{Takekoshi:1998, Xu:2003,
Thalhammer:2006}). Here, however, we find that the difference in
magnetic moments between the atoms and molecules can be made large
enough that the Stern-Gerlach effect due to the magnetic levitation
gradient can be used to separate atoms and molecules, allowing us to
produce pure samples of 2000 to 4000 RbCs molecules starting from
approximately 150000 Rb and 60000 Cs atoms \cite{Lercher:2011}. The
temperature of the molecular cloud is approximately the same as that of
the atomic sample, i.e. $100$ to $200$ nK.

We magnetoassociate at either the 197.06 or the 225.43~G resonance,
entering the bound-state manifold as seen in Fig.\ \ref{wiggle}. Below
each of these Feshbach resonances, there is a strongly avoided crossing
with the least-bound state, which we cannot jump over with our finite
magnetic switching capability. As a result, immediately after
magnetoassociation, the molecules transfer into the least-bound state
$|{-1}(1,3)s(1,3)\rangle$, which has a magnetic moment
$\mu=-1.3\mu_{\rm B}$, almost identical to that of the free atom pair.
In order to separate the atomic and molecular clouds, we switch off the
crossed optical dipole force trap confining the atom/molecule mixture
and quickly (in 0.5 ms) sweep $B$ down to the next avoided crossing,
below the 181.64 or 217.34~G resonance, respectively.

In the case of magnetoassociation at 197.06~G, we cross over onto the
low-field-seeking state $|{-6}(2,4)d(2,4)\rangle$ (with
$\mu\!=\!+2.0\mu_{\rm B}$) near 182~G, and then use another avoided
crossing (Fig.\ \ref{wiggle}, panel (a)) to transfer to the
high-field-seeking state $|{-2}(1,3)d(0,3)\rangle$ (with
$\mu=-0.9\mu_{\rm B}$). Just before we take the first of these two
crossovers, the magnetic field gradient is ramped up to a value
suitable for levitating the $|{-2}(1,3)d(0,3)\rangle$ molecules. At
this moment, the molecules are still in the least-bound state
$|{-1}(1,3)s(1,3)\rangle$ and are pushed upwards together with the
atoms. Rb $|1,1\rangle$ and Cs $|3,3\rangle$ have nearly the same
magnetic-moment-to-mass ratio at these field values and thus move
together. A large downward impulse is imparted to the molecules as
they pass through the low-field-seeking state. This separates the
atomic cloud from the molecular cloud. After going through the second
crossover, the molecules become high-field seekers that are levitated
exactly against gravity, and the optical dipole force trap is turned on
again, trapping the molecules. The Stern-Gerlach separation takes 3~ms
and produces a pure sample of up to 4000 molecules. These are observed
by a dissociation ramp backwards along the previous path, after which
the Rb and Cs atom clouds are imaged separately.

In the case of magnetoassociation at 225.43~G, we cross over near 217~G
onto the $|{-6}(2,4)d(2,2)\rangle$ state, which is also strongly
low-field-seeking.  To levitate the molecules in this state, the
direction of the current in the gradient coils must be switched,
causing a delay that results in additional atom-molecule collisions. In
this case, we produce pure clouds of typically 2000 molecules.

We note that the molecule creation efficiency of less than $10\%$ is
much lower than can be reached under optimized conditions for
single-species experiments, e.g.\ more than $25\%$ or even $30\%$
\cite{Mark2007siw,Knoop2008mfm,Mark2005eco} for the creation of
Feshbach molecules in a single-species BEC, and more than $90\%$
\cite{Danzl:ground:2010} for the creation of Feshbach molecules in the
two-atom shell of a single-species atomic Mott-insulator state. For the
present experiment we believe that we are limited by phase-space
density, which is of order unity for both clouds before they are
brought to overlap. We expect to increase the molecule creation
efficiency greatly once we are capable of overlapping the two atomic
samples in the quantum-degenerate regime in the presence of an optical
lattice, as discussed in Ref.~\cite{Lercher:2011}.

\subsection{Magnetic moment spectroscopy} \label{sec:magmom}

\begin{figure}[htbp]
\includegraphics[width=8.6cm] {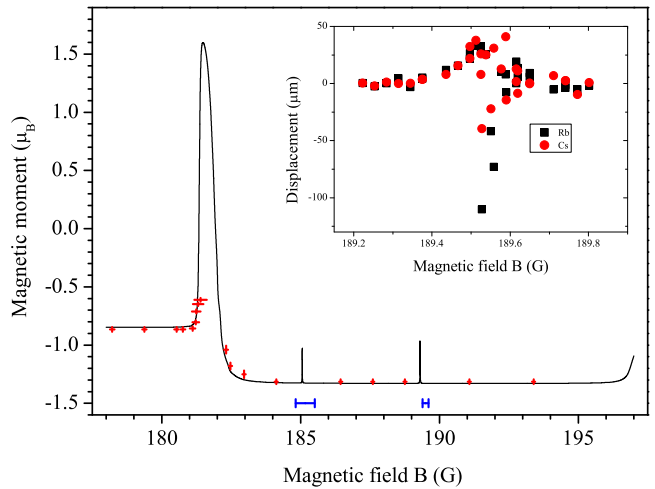}
\caption{[Color online.]
Calculated (solid line) and measured (red points) molecular magnetic
moment along the magnetoassociation route starting at the 197.06~G
Feshbach resonance. The two horizontal error bars mark the estimated
field values for two weak avoided crossings, which we detect by
variations in the cloud displacement at fixed gradient field; see
inset. The crossings are visible here as spikes in the calculated curve
and are indicated more explicitly in Figs.~\ref{wiggle}b and
~\ref{wiggle}c. Inset: Cloud displacement as a function of magnetic
field $B$ near the weak avoided crossing at $B=189.50\,$G. The strong
variation in the displacement data indicates the presence of the
crossing. Similar data is obtained for the weak avoided crossing near
$B=185.17\,$G.} \label{magmoment}
\end{figure}

We have measured the magnetic moments of the Feshbach molecules along
the high-field-seeking sections of the 197.06~G magnetoassociation
route. After re-trapping the pure $|{-2}(1,3)d(0,3)\rangle$ molecular
cloud, we backtrack to a magnetic field value $B$ where we are
interested in measuring the magnetic moment, and change the magnetic
field gradient. The dipole trap is then switched off and after
10--15~ms the molecules are dissociated and the fragments are imaged.
The field gradient that exactly levitates the molecules is scaled to
the field gradient needed to levitate Rb atoms at the same magnetic
field value. The Breit-Rabi equation is used to calculate the
Rb$|1,1\rangle$ magnetic moment at this field, and we multiply this by
the scaling factor (considering also the atomic and molecular masses)
to get the molecular magnetic moment. The measured magnetic moments
(Fig.~\ref{magmoment}) are consistent with those expected from the
coupled-channel calculations, which confirms our interpretation of the
197.06~G magnetoassociation route. The error in the magnetic moment is
dominated by the error in judging the correct levitation gradient due
to the large cloud sizes which result from expansion during the
levitation period. Since the experiment takes place in a field
gradient, the error in the magnetic field measurement is due mainly to
the difference in vertical position between the atomic Rb cloud used
for microwave-based magnetic field calibration and the molecular cloud.

%%%%%%%%%%%%%%%%%%%%%%%%%%%%%%%%%
%%%%%%%%%%%%%%%%%%%%%%%%%%%%%%%%%

Magnetic moment spectroscopy has also allowed us to estimate the
magnetic field values at which the two $|{-2}(1,3)d\rangle$ states
cross the least-bound state, as shown by arrows in Fig.~\ref{wiggle}.
Results for one of the crossings are shown in the inset to
Fig.~\ref{magmoment}. A very small increase in the magnetic moment is
seen near 189.50~G, which we interpret as a partial crossover onto
$|2(1,3)d(1,1)\rangle$ during the magnetic field sweep before
levitation; the corresponding avoided crossing is illustrated in
Fig.~\ref{wiggle}(c). We have tried to cross over to this state
adiabatically but have not been successful, most likely due to
technical magnetic field fluctuations. The
$|-(1,3)d(0,3)\rangle\leftrightarrow|{-1}(1,3)s(1,3)\rangle$
crossing, illustrated in Fig.~\ref{wiggle}(b), has also been observed
in this way. The magnetic moment signal produced by these crossings is
difficult to analyze because it is so weak, and the error bars shown in
Fig.~\ref{wiggle} simply span the range over which the magnetic moment
deviates from its background value.

\subsection{Bound-free modulation spectroscopy and binding energy of
the $|{-2}(1,3)d(0,3)\rangle$ state} \label{sec:13d03}

The binding energy of the $|{-2}(1,3)d(0,3)\rangle$ state proved to be
difficult to measure directly, presumably due to extremely weak
coupling to the atomic scattering channel. However, it was possible to
observe this state in the vicinity of the crossing with
$|{-6}(2,4)d(2,4)\rangle$, as shown in Fig.~\ref{wiggle}(a), using
bound-free magnetic-field modulation spectroscopy \cite{Regal2003cum}.
In this version of modulation spectroscopy, molecules that are produced
by magnetoassociation (as described in subsection \ref{FM}) are
dissociated when the energy $h f_\text{m}$ corresponding to the
modulation frequency $f_\text{m}$ is equal to or slightly greater than
the binding energy. The threshold frequency at which molecules begin to
be destroyed is associated with the binding energy. We observe the
bound-free transition by omitting from our experimental sequence the
reverse magnetoassociation ramp that is used to observe the molecules.
Any atoms that appear after applying the modulation are assumed to be
produced from molecule-atom transitions. Because the atomic signal
background is now very low, this method has inherent
signal-to-background advantages over free-bound spectroscopy, but the
low number of molecules increases the statistical noise. The
$|{-2}(1,3)d(0,3)\rangle$ state was observable only due to mixing with
$|{-6}(2,4)d(2,4)\rangle$ near the crossover at about 2.5 MHz$\times h$
binding energy. This is consistent with the fact that no Feshbach
resonance could be found for the $|{-2}(1,3)d(0,3)\rangle$ state near
its predicted intersection with the incoming scattering channel. Power
broadening causes the binding energy of the most deeply bound states to
be underestimated. While this effect was extrapolated to zero
intensity, the error bars shown in Fig.~\ref{wiggle} reflect our best
estimate of the possible systematic error that remains.

\section{Theory and Calculations}

The Hamiltonian for the interaction of two alkali-metal atoms may be written as
\begin{equation}
\frac{\hbar^2}{2\mu} \left[-R^{-1} \frac{d^2}{dR^2} R + \frac{\hat{L}^2}{R^2} \right] + \hat h_1 + \hat h_2 + \hat V(R),
\label{eq:SE}
\end{equation}
where $\mu$ is the reduced mass and $\hat L$ is the operator for the
end-over-end angular momentum of the two atoms about one another. The
monomer Hamiltonians including Zeeman terms are
\begin{equation}
\hat h_j = \zeta \hat \imath_j \cdot \hat s_j + g_e \mu_{\rm B}
B \, \hat s_{zj} + g_n \mu_{\rm B} B \, \hat \imath_{zj},
\label{eq:h-hat}
\end{equation}
where $\hat s_1$ and $\hat s_2$ represent the electron spins of the two
atoms and $\hat \imath_1$ and $\hat \imath_2$ represent the nuclear
spins. The constants $g_e$ and $g_n$ are the electron and nuclear
$g$-factors, $\mu_{\rm B}$ is the Bohr magneton, and $\hat s_z$ and
$\hat \imath_z$ represent the $z$-components of $\hat s$ and $\hat
\imath$ along a space-fixed $Z$ axis whose direction is defined by the
external magnetic field $B$. The atomic $g$-factors were taken from the
2006 CODATA adjustment of fundamental constants \cite{CODATA:2006} and
the $^{87}$Rb hyperfine constant from Bize \textit{et al.}\
\cite{Bize:1999}. The Cs hyperfine constant is exact by definition.

The interaction between the two atoms $\hat V(R)$ is
\begin{equation}
{\hat V}(R) = \hat V^{\rm c}(R) + \hat V^{\rm d}(R).
\label{eq:V-hat}
\end{equation}
Here $\hat V^{\rm c}(R)=V_0(R)\hat{\cal{P}}^{(0)} +
V_1(R)\hat{\cal{P}}^{(1)}$ is an isotropic potential operator that
depends on the potential energy curves $V_0(R)$ and $V_1(R)$ for the
respective $X{}^1\Sigma_g^+$ singlet and $a{}^3\Sigma_u^+$ triplet
states of the diatomic molecule. The singlet and triplet projectors
$\hat{\cal{P}}^{(0)}$ and $\hat{ \cal{P}}^{(1)}$ project onto subspaces
with total electron spin quantum numbers 0 and 1 respectively.
Figure~\ref{fig-curves} shows the two potential energy curves for RbCs.
The term $\hat V^{\rm d}(R)$ represents small, anisotropic
spin-dependent couplings, which are responsible for the avoided
crossings described in the experimental section and are discussed
further in Section~\ref{sec:secSO} below.

\subsection{Computational methods for bound states and scattering}

The three theoretical groups working on this problem used different
sets of computer codes that gave results in agreement with one another.
The methods used in Hannover to interpret the Fourier transform spectra
and Feshbach resonance positions are described in Ref.\
\cite{Pashov2007cot}. Those used at Temple University and NIST are
described in Ref.\ \cite{Tiesinga:na2:1998}. The methods used at Durham
are described below.

For the scattering and Feshbach bound states, we solve the
Schr\"odinger equation by coupled-channel methods, using a basis set
for the electron and nuclear spins in a fully decoupled representation,
\begin{equation}
|s_{\rm Rb} m_{s,{\rm Rb}}\rangle|i_{\rm Rb} m_{i,{\rm Rb}}\rangle
|s_{\rm Cs} m_{s,{\rm Cs}}\rangle |i_{\rm Cs} m_{i,{\rm Cs}}\rangle |L
M_L \rangle. \label{eqbasdecoup}
\end{equation}
The matrix elements of the different terms in the Hamiltonian in this
basis set are given in the Appendix of
Ref.~\cite{Hutson:Cs2-note:2008}. The calculations in this paper used
basis sets with all possible values of $m_s$ and $m_i$ for both atoms,
truncated at $L_{\rm max}=2$ unless otherwise indicated.

Scattering calculations are carried out using the MOLSCAT package
\cite{molscat:v14}, as modified to handle collisions in magnetic fields
\cite{Gonzalez-Martinez:2007}. At each magnetic field $B$, the
wavefunction log-derivative matrix at collision energy $E$ is
propagated from $R=0.3$ to 2.5 nm using the propagator of Manolopoulos
\cite{Manolopoulos:1986} with a fixed step size of 0.02 pm and from 2.5
to 1,500 nm using the Airy propagator \cite{Alexander:1987} with a
variable step size controlled by the parameter TOLHI=$10^{-5}$
\cite{Alexander:1984}. Scattering boundary conditions
\cite{Johnson:1973} are applied at $R=1,500$ nm to obtain the
scattering S-matrix. The energy-dependent $s$-wave scattering length
$a(k)$ is then obtained from the diagonal S-matrix element in the
incoming $L=0$ channel using the identity \cite{Hutson:res:2007}
\begin{equation}
\label{scat-length}
a(k) = \frac{1}{ik} \left(\frac{1-S_{00}}{1+S_{00}}\right),
\end{equation}
where $k^2=2\mu E/\hbar^2$. For $L=1$, this is generalized by replacing
$a$ with $a_1^3$ and $k$ with $k^3$.

Weakly bound levels for Feshbach molecules are obtained using a variant
of the propagation method described in
Ref.~\cite{Hutson:Cs2-note:2008}. The log-derivative matrix is
propagated outwards from $R=0.3$ to 2.5 nm with a fixed step size of
0.02 pm and inwards from 1,500 and 2.5 nm with a variable step size. In
Ref.~\cite{Hutson:Cs2-note:2008}, bound-state energies at a fixed value
of the magnetic field $B$ were located using the BOUND package
\cite{Hutson:bound:1993}, which converges on energies where the
smallest eigenvalue of the log-derivative matching determinant is zero
\cite{Hutson:CPC:1994}. However, for the purposes of the present work
we used a new package, FIELD, which instead works at fixed binding
energy and converges in a similar manner on the magnetic fields at
which bound states exist. BOUND and FIELD both implement a node-count
algorithm \cite{Hutson:CPC:1994} which makes it straightforward to
ensure that {\em all} bound states that exist in a particular range of
energy or field are located.

Zero-energy Feshbach resonances can in principle be located as fields
$B_{\rm res}$ at which the scattering length $a(B)$ passes through a
pole. However, with this method it is necessary first to search for
poles, and it is quite easy to miss narrow resonances. Since resonances
occur at fields where there is a bound state at zero energy, the FIELD
package provides a much cleaner approach: simply running FIELD at zero
energy provides a complete list of all fields at which zero-energy
Feshbach resonances exist.

\subsection{Representation of the potential curves}

The singlet and triplet curves are represented as described by Docenko
{\em et al.} \cite{Docenko:RbCs:2011}. In a central region from $R^{\rm
SR}_S$ to $R^{\rm LR}_S$, with $S=0$ or 1 for the singlet or triplet
state, respectively, the curves are well determined by the Fourier
transform spectra and are represented as finite power expansions of a
nonlinear function $\xi$ that depends on the internuclear separation
$R$,
\begin{equation}
V_S(R) = hc \sum_{i=0}^n a_i \xi^i(R),
\end{equation} where
\begin{equation}
\xi(R)= \frac{R-R_{\rm m}}{R+bR_{\rm m}}.
\end{equation}
The quantities $a_i$ and $b$ are fitting parameters, and $R_{\rm m}$ is
chosen to be near the equilibrium distance. At long range ($R
> R^{\rm LR}_S$), the potentials are
\begin{equation}
\begin{split}
V_S^{\rm LR}(R) = -C_6/R^6 - C_8/R^8 - C_{10}/R^{10}\\
-(-1)^S V_{\rm exch}(R),
\end{split}
\end{equation}
where the dispersion coefficients $C_n$ are common to both potentials.
The exchange contribution is \cite{Smirnov:1965}
\begin{equation}
V_{\rm exch}(R) = A_{\rm ex} (R/a_0)^\gamma \exp(-\beta R/a_0),
\end{equation}
and makes an attractive
contribution for the singlet and a repulsive contribution for the
triplet. $\beta$ and $\gamma$ are related via $\gamma=7/\beta-1$ and
are obtained from the ionization energies
of Rb and Cs \cite{Smirnov:1965}, and $A_{\rm ex}$ is a fitting
parameter. The mid-range potentials are constrained to match the
long-range potentials at $R_S^{\rm LR}$. Lastly, the potentials are
extended to short range ($R < R_S^{\rm SR}$) with simple repulsive
terms,
\begin{equation}
V_S^{\rm SR}(R) = A^{\rm SR}_S + B^{\rm SR}_S [a_0/R]^{N},
\end{equation}
where $A^{\rm SR}_S=V_S(R_S^{\rm SR})-B^{\rm SR}_S [a_0/R_S^{\rm
SR}]^{N}$ is chosen to match the short-range and mid-range potentials
at $R_S^{\rm SR}$.

\subsection{Magnetic dipole interaction and second-order spin-orbit
coupling} \label{sec:secSO}

At long range, the coupling $\hat V^{\rm d}(R)$ of Eq.\ \ref{eq:V-hat}
has a simple magnetic dipole-dipole form that varies as
$1/R^3$~\cite{Stoof:1988, Moerdijk:1995}. However, for heavy atoms it
is known that second-order spin-orbit coupling provides an additional
contribution that has the same tensor form as the dipole-dipole term
and dominates at short range \cite{Mies:1996, Kotochigova:2001}. In the
present work, $\hat V^{\rm d}(R)$ is represented as
\begin{equation}
\label{eq:Vd} \hat V^{\rm d}(R) = \lambda(R) \left ( \hat s_1\cdot
\hat s_2 -3 (\hat s_1 \cdot \vec e_R)(\hat s_2 \cdot \vec e_R)
\right ) \,,
\end{equation}
where $\vec e_R$ is a unit vector along the internuclear axis and
$\lambda$ is an $R$-dependent coupling constant. This term couples the
electron spins of Rb and Cs atoms to the molecular axis. In particular,
it couples the even partial waves ($s$, $d$, ...) with one another and
does the same for the odd partial waves ($p$, $f$, ...).

In the present work the second-order term was evaluated from electronic
structure calculations in a manner similar to that described in
Ref.~\cite{Kotochigova:2001}, using a relativistic
configuration-interaction valence bond (RCI-VB) method. The molecular
wave function is constructed from atomic orbitals localized at the
different atomic centers. Configuration interaction (CI) coefficients
are obtained by solving a generalized eigenvalue matrix problem of the
relativistic electronic Hamiltonian based on a nonorthogonal basis set.
At short internuclear separations, the one-electron orbitals from
different centers have considerable overlap or nonorthogonality, which
gives rise to a large exchange interaction and thereby creates the
bond. For large internuclear separations, the molecular wave function
automatically obtains a pure atomic form, which is the correct
asymptotic limit for any molecular wave function. In our version of the
RCI-VB method, the atomic Slater determinants are constructed from
one-electron numerical Dirac-Fock functions for occupied core and
valence orbitals and numerical Sturmian functions for virtual or
unoccupied orbitals. These Sturmian orbitals are obtained by solving
integro-differential Dirac-Fock-Sturm equations
\cite{Kotochigova:2005}.

For RbCs, all occupied orbitals up to the 4s$^2$ shell in Rb and the
5s$^2$ shell in Cs are defined as the core orbitals. The 4p$^6$
orbitals in Rb and 5p$^6$ orbitals in Cs are included in the
core-valence subspace, allowing single and double excitations. The 5s,
5p, 4d, 6s, and 6p orbitals of Rb and 6s, 6p, 5d, 7s, and 7p orbitals
of Cs are added to the active subspace with single, double, and triple
occupancy. In addition, we included virtual Sturm 5d, 4f, 7s, and 7p
orbitals of Rb and 6d, 4f, 8s, and 8p orbitals of Cs to complete the
active space. Up to double occupancy is allowed for these virtual
orbitals.

\begin{figure}
\includegraphics[scale=0.4]{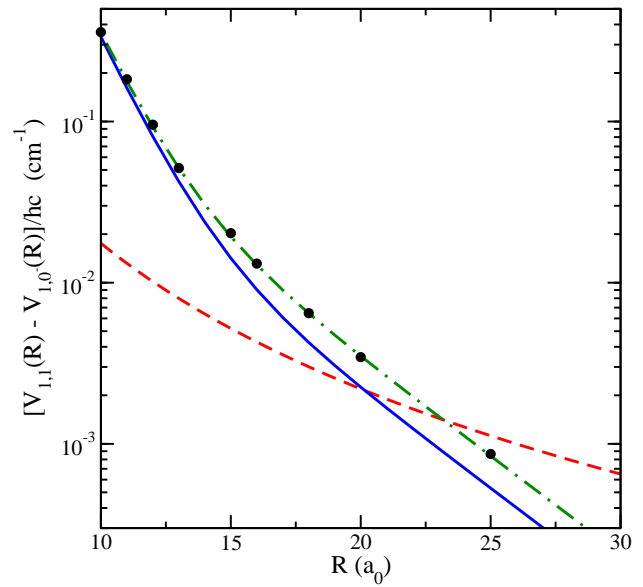}
\caption{[Color online] Study of the second-order spin-orbit
interaction energy $V_{1,1}(R)-V_{1,0^-}(R)$ as a function of
internuclear separation $R$. The filled circles are the result of the
{\it ab initio} RCI-VB electronic structure calculation. The green
dash-dotted is a fit to the {\it ab initio} values using the functional
form of Eq. \ref{lambdaR}. The blue line corresponds to the
second-order spin-orbit interaction energy optimized to reproduce the
location of the observed magnetic Feshbach resonances for A$_{\rm
2SO}^{\rm long}$ = -0.03310. For comparison the absolute value of the
corresponding splitting due to the magnetic dipole-dipole interaction
is shown by a red dashed line.} \label{2nd}
\end{figure}

Our relativistic valence bond method calculates the second-order
spin-orbit splitting nonperturbatively. The calculation finds the
energetically lowest $\Omega=0^-$ and $1$ states, which correspond to
the two fine-structure components of the $S=1$ $a{}^3\Sigma^+$
Born-Oppenheimer potential. We denote the relativistic potentials by
$V_{S,\Omega}(R)$. The difference
$V_{1,1}(R)-V_{1,0^-}(R)=-(3/2)\lambda(R)$ provides the second-order
spin-orbit splitting shown in Fig.~\ref{2nd}. Also shown is the
strength of the spin-spin dipole interaction, which leads to a
splitting between the $0^-$ and $1$ Born-Oppenheimer potentials with
opposite sign compared to the second order spin-orbit contribution.

The second-order spin-orbit splitting has a nearly exponential
dependence on $R$ and lies about half-way between the values for Rb$_2$
and Cs$_2$ molecules calculated previously \cite{Kotochigova:2001}. The
results of the electronic structure calculations were fitted to a
biexponential form, so that the overall form of $\lambda(R)$ is
\begin{eqnarray}
\label{eq:lambda}
\lambda(R) &\!\!=&\!\! E_{\rm h} \alpha^2 \bigg[
A_{\rm 2SO}^{\rm short} \exp\left(-\beta_{\rm 2SO}^{\rm short}(R/a_0)\right)
\nonumber\\
&\!\!+&\!\!\!\! A_{\rm 2SO}^{\rm long}
\exp\left(-\beta_{\rm 2SO}^{\rm long}(R/a_0)\right)
+  \frac{1}{(R/a_0)^3}\bigg],
\label{lambdaR}
\end{eqnarray}
where $\alpha\approx 1/137$ is the atomic fine-structure constant. The
parameters obtained from fitting to the electronic structure
calculations are $A_{\rm 2SO}^{\rm short} = -50.974$, $A_{\rm 2SO}^{\rm
long} = -0.0525$, $\beta_{\rm 2SO}^{\rm short} = 0.80$ and $\beta_{\rm
2SO}^{\rm long} = 0.28$.  However, in fitting to the weakly bound
levels, this coupling function was found to be too strong to reproduce
the avoided crossings shown in Fig.~\ref{wiggle}. We therefore retained
the functional form (\ref{eq:lambda}) but allowed the parameter $A_{\rm
2SO}^{\rm long}$ to vary in the least-squares fit to the experimental
results described below.

\subsection{Assignment of quantum numbers}

At the start of this work, the singlet and triplet scattering lengths
$a_S$ and $a_T$ for RbCs were unknown within wide ranges and there was
no assignment of quantum numbers to the Feshbach resonances of
Ref.~\cite{Pilch:2009}. However, the identification of a bound state in
the $|1,1\rangle + |3,3\rangle$ channel bound by only about 110 kHz
placed the possible values of $a_S$ and $a_T$ along a well-defined
curve in the upper-right quadrant of $a_S,a_T$ space. We therefore used a
pre-publication version of the mid-range RbCs potentials of Docenko
{\em et al.}\ \cite{Docenko:RbCs:2011}, modified to allow us to vary
the scattering lengths, and carried out coupled-channel calculations at
a number of points along this line to identify lists of $s$-wave
Feshbach resonances. By altering the long-range coefficients and
inner-wall parameters of this potential, we were able to produce a
resonance pattern that approximately matched the experimental one, and
also gave a pattern of bound states similar to that from free-bound
spectroscopy. A key feature that strengthened our confidence in this
assignment was that it predicted two very weak crossings between the
least-bound state near 110 kHz and two $n=-2$ states, as shown in Fig.\
\ref{wiggle}. The presence of these crossings was then confirmed by
experiment as described in Section \ref{sec:magmom} above.

At around this time, the final version of the spectroscopic potentials
of Ref.~\cite{Docenko:RbCs:2011} became available.  These had different
numbers of singlet and triplet bound states from the preliminary
version, but approximately the correct scattering lengths. We therefore
used this potential to produce resonance patterns and a bound state map
of the region immediately below the lowest threshold. This gave a good
match to the experimentally observed Feshbach resonance positions, but
placed the two $n=-2$ states that cross the least-bound state between
180~G and 190~G at fields about 3 G too low. In addition, the avoided
crossings between the $n=-6$ states and the least-bound state were
broader than was found experimentally. We therefore embarked on a
two-part least-squares refinement, beginning from the potential of
Ref.\ \cite{Docenko:RbCs:2011}, as described below.

\subsection{Least-squares refinement}

The Feshbach bound states and resonance positions are strongly
sensitive to the long-range potential and to the scattering lengths,
but only weakly sensitive to the details of the potential in the well
region. The Fourier Transform spectra, by contrast, are very sensitive
to the well region. The potentials are determined in an iterative loop
using the data sequentially, as was successfully applied for example in
Ref.\ \cite{Pashov2007cot}. First, in the least-squares fit to weakly
bound states and Feshbach resonance positions, the potential curves in
the central region were held fixed but the long-range coefficients
$C_6$ and $C_8$ were allowed to vary. In addition, the parameters
$B_0^{\rm SR}$ and $B_1^{\rm SR}$ were varied, allowing the inner walls
of the two potential curves to move sufficiently to adjust the singlet
and triplet scattering lengths independently of $C_6$ and $C_8$. The
scaling factor for the long-range part of the 2nd-order spin-orbit
coupling was also varied in this step. In the second step, the
long-range function was held fixed and the inner parts of the
potentials were varied to fit the large set of results from Fourier
Transform spectroscopy, adding as data the uncoupled last-bound levels
constructed from the fit in the first step. Two iterations were
sufficient to achieve convergence between the two different
least-squares procedures.

The propagator approach to locating bound states and resonances,
implemented in the BOUND and FIELD programs, is fast enough to be
incorporated in a least-squares fitting program. Nevertheless, it is
still slow enough that these calculations form the major time-consuming
step in a least-squares refinement procedure.  Furthermore, the
parameter set used is highly correlated. Under these circumstances, a
fully automated approach to fitting is unreliable: individual
least-squares steps often reach points in parameter space where the
levels have moved too far to be identified reliably, particularly in
the early stages of fitting. We therefore carried out this stage of the
fitting using the I-NoLLS package \cite{I-NoLLS} (Interactive
Non-Linear Least-Squares), which gives the user interactive control
over step lengths and assignments as the fit proceeds. This allowed us
to converge on a minimum in the sum of weighted squares in a relatively
small number of steps.

The measurements on weakly bound states described above complement the
measurements of the positions of Feshbach resonances. In particular:
(i) the position of the least-bound state is sensitive to the
background scattering length in the incoming $|1,1\rangle$ +
$|3,3\rangle$ channel; (ii) the strengths of the avoided crossings
between the least-bound state and the ramping $n=-6$ states from the
(2,4) threshold are sensitive to the magnitude of the 2nd-order
spin-orbit coupling; (iii) the positions of the $n=-2$ states
associated with the (1,3) threshold, observed through their avoided
crossings with the least-bound state, are sensitive to the long-range
$C_6$ coefficient, but relatively uncontaminated by the influence of
$C_8$, which becomes important for deeper levels. In combination with
the Feshbach resonances due to $n=-6$ states, whose position is
significantly influenced by the $C_8$ coefficient, the $n=-2$ levels
open the way for $C_6$ and $C_8$ to be determined separately.

\begin{table}[htpb]
\caption{Quality of fit to Feshbach bound states and resonance
positions. Top section: a complete list of the fields (in G) for all
calculated $s$-wave resonances in the region 10 to 560~G using an $sd$
basis, together with quantum labels as explained in the text. Some
calculated resonances have not been observed experimentally. Center
section: fields used to characterize the ramping states
$|-6(2,4)d(2,4)\rangle$ and $|-6(2,4)d(2,3)\rangle$ and their avoided
crossings with the least-bound state. Bottom section: the binding
energy of the $|-2(1,3)d(0,3)\rangle$ bound state at 181.18 G, just
below its crossing with $|-6(2,4)d(2,4)\rangle$.  For all states here,
the total angular projection quantum number $M$ is 4. The uncertainties
quoted here are those that define the weights used in our least-squares
fit.} \label{quality-of-fit}

\begin{tabular}{rcccrclcl}
\hline\hline
\omit\hfill\vrule height 2ex depth 1ex width 0pt $B_{\rm calc}$\hfill&\qquad& $B_{\rm obs}$&\multispan3\hfill$B_{\rm obs}-B_{\rm calc}$\hfill & Unc. &\qquad& quantum labels\\
\hline
 87.25 &&        &&         &&\vrule height 3ex width 0pt &&$\left|-2\,(1,3)\,d\,(-1,3)\right\rangle$\\
123.09 &&        &&         &&&&$\left|-2\,(1,3)\,d\,(0,2)\right\rangle$\\
181.63 && 181.64 &&  $0.01$& & 0.10  &&$\left|-6\,(2,4)\,d\,(2,4)\right\rangle$\\
197.07 && 197.06 && $-0.01$& & 0.046 &&$\left|-6\,(2,4)\,d\,(2,3)\right\rangle$\\
217.33 && 217.34 &&  $0.01$&  & 0.047 &&$\left|-6\,(2,4)\,d\,(2,2)\right\rangle$\\
225.47 && 225.43 && $-0.04$& & 0.034 &&$\left|-6\,(2,4)\,d\,(1,4)\right\rangle$\\
242.25 && 242.29 &&   0.04&  & 0.047 &&$\left|-6\,(2,4)\,d\,(2,1)\right\rangle$\\
247.28 && 247.32 &&   0.04&  & 0.048 &&$\left|-6\,(2,4)\,d\,(1,3)\right\rangle$\\
272.81 && 272.80 && $-0.01$& & 0.043 &&$\left|-6\,(2,4)\,d\,(2,0)\right\rangle$\\
       && 273.45 &&         &&0.04&\\
273.69 && 273.76 &&  0.07&   & 0.043 &&$\left|-6\,(2,4)\,d\,(1,2)\right\rangle$\\
279.02 && 279.12 &&  0.10&   & 0.048 &&$\left|-6\,(2,4)\,s\,(2,2)\right\rangle$\\
286.68 && 286.76 &&  0.08&   & 0.047 &&$\left|-6\,(2,4)\,d\,(0,4)\right\rangle$\\
308.45 && 308.44 && $-0.01$& & 0.045 &&$\left|-6\,(2,4)\,d\,(1,1)\right\rangle$\\
310.71 && 310.69 && $-0.02$&  & 0.056 &&$\left|-6\,(2,4)\,s\,(1,3)\right\rangle$\\
314.56 && 314.74 &&  0.18&   & 0.11  &&$\left|-6\,(2,4)\,d\,(0,3)\right\rangle$\\
352.74 && 352.65 && $-0.09$& & 0.34  &&$\left|-6\,(2,4)\,s\,(0,4)\right\rangle$\\
353.57 &&        &&        & &&&$\left|-6\,(2,4)\,d\,(0,2)\right\rangle$\\
381.28 && 381.34 &&  0.06&   & 0.047 &&$\left|-6\,(2,4)\,d\,(-1,4)\right\rangle$\\
408.63 &&        &&       &  &&&$\left|-2\,(1,3)\,d\,(1,2)\right\rangle$\\
422.04 && 421.93 && $-0.11$& & 0.047 &&$\left|-6\,(2,4)\,d\,(-1,3)\right\rangle$\\
552.75 &&        &&         &&&&$\left|-6\,(2,4)\,d\,(-2,4)\right\rangle$\\
\hline
 185.24 && 185.34$^*$ && 0.10& & 0.35&&$\left|-2\,(1,3)\,d\,(0,3)\right\rangle$\\
189.47 && 189.66$^\dagger$ &&  0.19& & 0.10&&$\left|-2\,(1,3)\,d\,(1,1)\right\rangle$\\
\end{tabular}

\begin{tabular}{lrrrl}
\hline\vspace{2pt}
&\omit\hfill $B_{\rm calc}$\hfill&\omit\hfill$B_{\rm obs}$\hfill
&
\vrule height 2ex width 0pt
$B_{\rm obs}-B_{\rm calc}$&Unc.\\
\hline\vspace{2pt}
$B_{(2,4)}$ at $-1.02$ MHz & 181.729 & 181.758 & 0.030 \quad\qquad& 0.03\\
$B_{(2,3)}$ at $-0.84$ MHz & 196.978 & 196.946 & $-0.019$ \quad\qquad& 0.02\\
\vspace{2pt}
$B^+_{(2,4)}$ at $-0.030$ MHz & 181.381 & 181.380 & \\
$B^-_{(2,4)}$ at $-0.210$ MHz & 182.358 & 182.316 & $-0.042$ \quad\qquad& 0.03\\
$B^-_{(2,4)}-B^+_{(2,4)}$ & 0.977 & 0.936 & $-0.041$ \quad\qquad& 0.07\\
$B^+_{(2,3)}$ at $-0.030$ MHz & 196.991 & 196.950 & \\
$B^-_{(2,3)}$ at $-0.185$ MHz & 197.300 & 197.278 & $-0.022$ \quad\qquad& 0.03\\
$B^-_{(2,3)}-B^+_{(2,3)}$ & 0.309 & 0.328 & 0.019 \quad\qquad& 0.06\\
\hline\vspace{2pt}
&\omit\vrule height 2ex width 0pt \hfill $-E_{\rm calc}$\hfill&\omit\hfill$-E_{\rm obs}$\hfill
&
$-E_{\rm obs}+E_{\rm calc}$&Unc.\\
\hline
\vrule height 2.5ex width 0pt
$E^{-2}_{(0,3)}$ at 181.18 G& 2.767 & 2.525 & $-0.242$ \quad\qquad& 0.10\\
(MHz)\\
\hline\hline
\end{tabular}
\linebreak\hspace*{-0.4cm} $^*$ Resonance position extrapolated from
avoided crossing at 185.17~G.\linebreak\hspace*{-0.4cm} $^\dagger$
Resonance position extrapolated from avoided crossing at 189.50~G.
\end{table}

Once we were confident of the assignment of the weakly bound states and
Feshbach resonances, we therefore carried out least-squares refinement
of the potential using the I-NoLLS package in the 5-parameter space
$B_0^{\rm SR}$, $B_1^{\rm SR}$, $C_6$, $C_8$, $A_{\rm 2SO}^{\rm long}$.
The set of experimental results used for this stage of fitting is
listed in Table \ref{quality-of-fit}. It consists of the magnetic
fields for all the measured $s$-wave resonances, except the resonance
at 273.45 G, which we attribute to a bound state of $g$ character, and
is supplemented by a selection from the measurements of the binding
energies: (i) two additional resonance positions for the $n=-2$ states,
obtained from the positions of the avoided crossings between the $n=-2$
states and the least-bound state by a (very short) extrapolation to
zero energy using the calculated slopes of the $n=-2$ states; (ii)
fields at which the bound states $|{-6}(2,4)d(2,4)\rangle$ and
$|{-6}(2,4)d(2,3)\rangle$ exist near 1 MHz; (iii) four fields at which
bound states exist near 110 kHz, designated $B^+_{(m_{\rm Rb},m_{\rm
Cs})}$ and $B^-_{(m_{\rm Rb},m_{\rm Cs})}$, just above and just below
the avoided crossings between the least-bound state and the
$|{-6}(2,4)d(2,4)\rangle$ and $|{-6}(2,4)d(2,3)\rangle$ states; to
improve the determination of the 2nd-order spin-orbit coupling, two of
these were included as field {\em differences} $B^--B^+$ between levels
just above and just below each crossing; (iv) the energy of the
$|{-2}(1,3)d(0,3)\rangle$ state at 181.18~G, just below its crossing
with $|{-6}(2,4)d(2,4)\rangle$. The quantity optimized in the
least-squares fits was the sum of squares of residuals ((observed $-$
calculated)/uncertainty), with the uncertainties listed in Table
\ref{quality-of-fit}.

\begin{table}[htpb]
\caption{Potential parameters and derived quantities resulting from
least-squares fitting to Feshbach bound states and resonance
positions.} \label{fitparms}
\begin{tabular}{l|@{\quad}r@{}l@{\quad}r@{}lcl}
\hline\hline
&\multispan2 \hfill fitted value\hfill & \multispan2\vrule height 2.5ex width 0pt 95\% confidence \hfill& \qquad & sensitivity \\
&\multispan2              & \multispan2 limit\hfill           &             \\
\hline
\vrule height 3ex width 0pt $B_0^{\rm SR}$ ($E_{\rm h}$) & 6960&.7   &710& && 0.5  \\
$B_1^{\rm SR}$ ($E_{\rm h}$) & 19793&.3 &110&  && 0.1  \\
$A^{\rm long}_{\rm 2SO}$ & $-0$&.0331  & 0&.0028   && 0.0001\\
$C_6$ ($E_{\rm h}a_0^6$) & 5693&.7056 & 2&.2  && 0.0004\\
$C_8$ ($E_{\rm h}a_0^8$) & 796487&.36  &1900&  && 0.3\\
\hline
\vrule height 2.5ex width 0pt derived &\multispan2\hfill value\hfill& \multispan2 \hfill uncertainty\hfill \\ parameters\\
\hline
$a_S$ ($a_0$) &997& &11\\
$a_T$ ($a_0$) & 513&.3 & 2&.2\\
\hline\hline
\end{tabular}
\end{table}

\begin{table}
\fontsize{8pt}{13pt}\selectfont \caption{Parameters of the analytic
representation of the potential of state \Xstate. The energy reference
is the dissociation asymptote.}
\label{singlet}   % pot used 122
\begin{tabular}{c c}%{@{\extracolsep{\fill}}|lr|}
\hline
   & $R < R_\mathrm{SR}=$ 0.3315 nm    \\
\hline
   $A^{\mathrm{SR}\ast}_0/hc$ & -0.407634031$\times 10^{4}$ \wn \\
   $B^{\mathrm{SR}}_0/hc$ &1.52770630$\times 10^{9}$  \wn\\
   $N_0$    & 7 \\
\hline
   &$R_\mathrm{SR} \leq R \leq R_\mathrm{LR}=$ 1.150 nm    \\
\hline
    $b$ &   $ 0.09$              \\
    $R_\mathrm{m}$ & 0.442708150 nm               \\
    $a_{0}$ &  -3836.36509 \wn\\
    $a_{1}$ &-0.0369980716645394794$ $ \wn\\
    $a_{2}$ & 0.447519742785341805$\times 10^{ 5}$ \wn\\
    $a_{3}$ &-0.134065881674135253$\times 10^{ 5}$ \wn\\
    $a_{4}$ &-0.112246913875781145$\times 10^{ 6}$ \wn\\
    $a_{5}$ &-0.680373468487243954$\times 10^{ 5}$ \wn\\
    $a_{6}$ & 0.124395856928352383$\times 10^{ 6}$ \wn\\
    $a_{7}$ &-0.527808915105630062$\times 10^{ 6}$ \wn\\
    $a_{8}$ & 0.160604050855185674$\times 10^{ 7}$ \wn\\
    $a_{9}$ & 0.856669313055434823$\times 10^{ 7}$ \wn\\
   $a_{10}$ &-0.423220682973604128$\times 10^{ 8}$ \wn\\
   $a_{11}$ &-0.846286860630152822$\times 10^{ 8}$ \wn\\
   $a_{12}$ & 0.775110557475278497$\times 10^{ 9}$ \wn\\
   $a_{13}$ & 0.208102060193851382$\times 10^{ 9}$ \wn\\
   $a_{14}$ &-0.762262944271048737$\times 10^{10}$ \wn\\
   $a_{15}$ & 0.645280096247728157$\times 10^{10}$ \wn\\
   $a_{16}$ & 0.358089708848128967$\times 10^{11}$ \wn\\
   $a_{17}$ &-0.685156406423631516$\times 10^{11}$ \wn\\
   $a_{18}$ &-0.340359743040435295$\times 10^{11}$ \wn\\
   $a_{19}$ & 0.204117122912590576$\times 10^{12}$ \wn\\
   $a_{20}$ &-0.207876500106921722$\times 10^{12}$ \wn\\
   $a_{21}$ & 0.712777331768994293$\times 10^{11}$ \wn\\
\hline
   & $ R > R_\mathrm{LR}$\\
\hline
 $C_6$  & 5693.7056 $E_{\rm h}a_0^6$\\
 $C_8$  & 796487.36 $E_{\rm h}a_0^8$\\
 $C_{10}$ &95332817 $E_{\rm h}a_0^{10}$\\
 ${A_\mathrm{ex}}/hc$ &0.37664685$\times 10^{3}$ \wn \\
 ${\gamma}$ &5.427916    \\
 ${\beta}$ &1.0890  \\
\hline
\end{tabular}
\linebreak\hspace*{-0.4cm}$^*$ This parameter is set to give continuity
between the short-range and mid-range functional forms.
\end{table}

\begin{table}
\fontsize{8pt}{13pt}\selectfont \caption{Parameters of the analytic
representation of the potential of state \astate. The energy reference
is the dissociation asymptote.}
\label{triplet}    % pot used 122
\begin{tabular}{c c}  %{@{\extracolsep{\fill}}|lr|}
\hline
   &  $R < R_\mathrm{SR}=$ 0.522 nm    \\
\hline
   $A^{\mathrm{SR}\ast}_1/hc$ & -0.500680370$\times 10^{3}$ \wn \\
   $B^{\mathrm{SR}}_1/hc$ & 4.34413885$\times 10^{9}$  \wn\\
   $N_1$    & 7 \\
\hline
   & $R_\mathrm{SR} \leq R \leq R_\mathrm{LR}=$ 1.200 nm    \\
\hline
    $b$ &   $ 0.06$              \\
    $R_\mathrm{m}$ & 0.62193776 nm               \\
    $a_{0}$ &  -259.33587 \wn\\
    $a_{1}$ & 0.1466188573699344914$ $ \wn\\
    $a_{2}$ & 0.525743927693154455$\times 10^{ 4}$ \wn\\
    $a_{3}$ &-0.122790966318838728$\times 10^{ 5}$ \wn\\
    $a_{4}$ & 0.175565797136193828$\times 10^{ 4}$ \wn\\
    $a_{5}$ & 0.173795490253058379$\times 10^{ 5}$ \wn\\
    $a_{6}$ &-0.119112720845007316$\times 10^{ 5}$ \wn\\
    $a_{7}$ &-0.245659148870101490$\times 10^{ 5}$ \wn\\
    $a_{8}$ & 0.303380094883701415$\times 10^{ 6}$ \wn\\
    $a_{9}$ &-0.100054913157079869$\times 10^{ 7}$ \wn\\
   $a_{10}$ &-0.296340813141656632$\times 10^{ 6}$ \wn\\
   $a_{11}$ & 0.997302450614721887$\times 10^{ 7}$ \wn\\
   $a_{12}$ &-0.272673123492070958$\times 10^{ 8}$ \wn\\
   $a_{13}$ & 0.323269132716538832$\times 10^{ 8}$ \wn\\
   $a_{14}$ &-0.147953587185832486$\times 10^{ 8}$ \wn\\
\hline
    & $ R > R_\mathrm{LR}$ \\
\hline
 $C_6$  & 5693.7056 $E_{\rm h}a_0^6$\\
 $C_8$  & 796487.36 $E_{\rm h}a_0^8$\\
 $C_{10}$ &95332817 $E_{\rm h}a_0^{10}$\\
 ${A_\mathrm{ex}}/hc$ &0.37664685$\times 10^{3}$ \wn \\
 ${\gamma}$ &5.427916    \\
 ${\beta}$ &1.0890  \\
\hline
\end{tabular}
\linebreak\hspace*{-0.4cm}$^*$ This parameter is set to give continuity
between the short-range and mid-range functional forms.
\end{table}

\subsection{Final potential}\label{sec:final}

At the conclusion of the two-part least-squares refinement procedure
described above, we arrived at the potentials given in Tables
\ref{fitparms}, \ref{singlet} and \ref{triplet}. Table \ref{fitparms}
gives the parameters from fitting to the Feshbach bound states and
resonance positions, whereas Tables \ref{singlet} and \ref{triplet}
give the full potentials for the singlet and triplet states,
respectively.

Although the Feshbach bound states and resonance positions do allow all
5 parameters in Table \ref{fitparms} to be extracted, they are very
highly correlated. The table therefore gives both 95\% confidence
limits and parameter sensitivities as defined by Le~Roy
\cite{LeRoy:1998}. The 95\% confidence limits are correlated properties
that describe the uncertainty in an individual parameter, but the
parameters need to be specified to within their sensitivities, not
their confidence limits, in order to reproduce the results of the
calculations.

The new version of the electronic potentials for the RbCs ground-state
system reproduces the Fourier transform spectra as accurately as the
original version of Ref.~\cite{Docenko:RbCs:2011}, with the important
improvement that it can also accurately reproduce properties relating
to the very top of the electronic potentials, such as Feshbach spectra.
There are some remaining deviations between the observed and calculated
positions of the $n=-2$ states as shown in the lower panels of Figure
\ref{wiggle}, but in view of the possible systematic errors in the
corresponding measurements described in Sections \ref{sec:13d03} and
\ref{sec:magmom} above, these are not a great cause for concern.

\begin{figure*}[tbp]
\includegraphics[angle=-90,width=1.96\columnwidth]{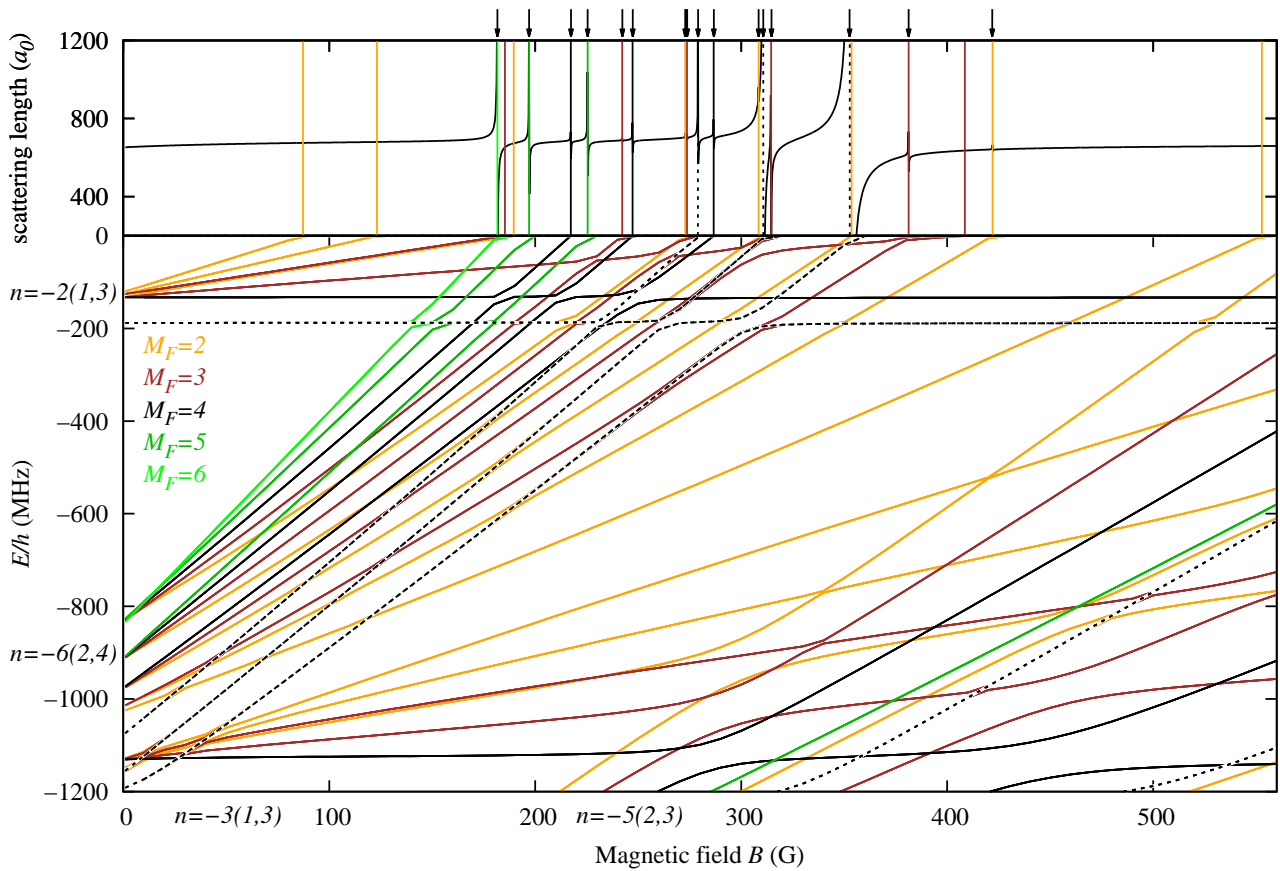}
\caption{[Color online.] Weakly bound states of RbCs for $M=4$
at fields up to 550~G, calculated using the final fitted potentials,
together with the scattering length at the $|1,1\rangle + |3,3\rangle$
threshold, calculated at $E=160$ nK. Bound states are plotted in a
colour corresponding to their value of $M_F$, as shown on the figure.
States arising from $L=2$ ($d$ states) are shown as solid lines, and
from $L=0$ ($s$ states) are shown as dashed lines. The resonance
positions are marked on the scattering length plot as vertical lines
with the same color as the bound state that they arise from. The slanted text on the left-hand axis and below the bottom axis indicates which vibrational and hyperfine manifold the $L=2$ bound states arise from. The least-bound state $|{-1}(1,3)s(1,3)\rangle$ is within the thickness of the zero line on this scale. The observed positions of incoming $s$-wave
resonances are shown as arrows above the plot.} \label{s-wave-low}
\end{figure*}

The final results for the resonance positions and weakly bound states
are listed in Table \ref{quality-of-fit}, together with the quality of
fit to the experimental data and the quantum label assignments. The
calculated $s$-wave scattering length and its match to the resonance
positions is shown in Figure \ref{s-wave-low}, together with an
overview of the bound states responsible for the resonances.

The singlet and triplet scattering lengths $a_S$ and $a_T$ obtained
from the fitted potentials are included in Table \ref{fitparms},
together with their fully correlated uncertainties, calculated as
described in Ref.~\cite{LeRoy:1998}. The background scattering length
derived for the $|1,1\rangle + |3,3\rangle$ channel is $651\pm10\ a_0$,
calculated at $B=500$~G far from resonances. We have also calculated
the binding energy of the least-bound state for $L=0$ at $B=211$~G
(this value was chosen to represent a value far from resonance within
the region for which experimental values are available) and found it to
be $110\pm2$ kHz $\times h$.

\subsection{Independent tests and predictions}

\subsubsection{Resonances in $p$-wave scattering}\label{sec:p-wave}

\begin{figure}[tbp]
\includegraphics[angle=-90,width=\columnwidth]{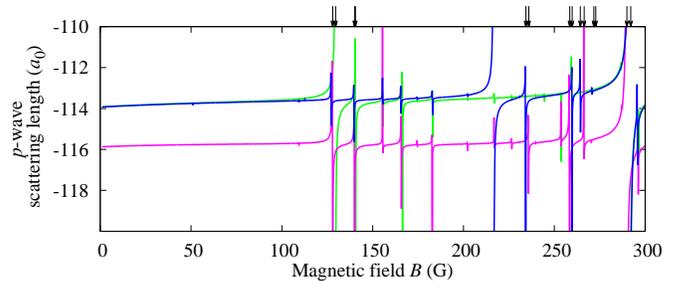}
\caption{[Color online.] The $p$-wave scattering lengths at the
$|1,1\rangle + |3,3\rangle$ threshold for the three values of $M$
allowed for $p$-wave scattering of Rb + Cs, calculated at $E=7\ \mu$K using the final fitted potentials. Results for $M_F=3$, 4, and 5 are shown as blue (dark gray), magenta (gray), and green (light gray), respectively. The observed positions of $p$-wave resonances are shown as arrows at the top of the graph.}
\label{p-wave}
\end{figure}

As described above, some of the resonances observed by Pilch {\em et
al.}\ \cite{Pilch:2009} do not appear for the Rb+Cs mixture at the
lower temperatures studied in the present work and are assigned as
resonances in $p$-wave scattering. When $L_\text{c}=1$, $M_{L_\text{c}}$ can take values $-1$, 0 or +1, so $M$ can be 3, 4, or 5 at the $|1,1\rangle +
|3,3\rangle$ threshold with $M_F=4$. Figure \ref{p-wave} compares the
observed $p$-wave resonance positions with the $p$-wave scattering
lengths for these three values of $M$, calculated using the fitted
potentials. It may be seen that the observed resonances correspond
quite well to a subset of the calculated resonances, although it is not
altogether clear why Pilch {\em et al.}\ \cite{Pilch:2009} observed
some $p$-wave resonances and not others.

\subsubsection{Resonances at the $|2,-1\rangle + |3,3\rangle$
threshold}\label{sec:excited}

\begin{figure}[tbp]
\includegraphics[angle=-90,width=\columnwidth]{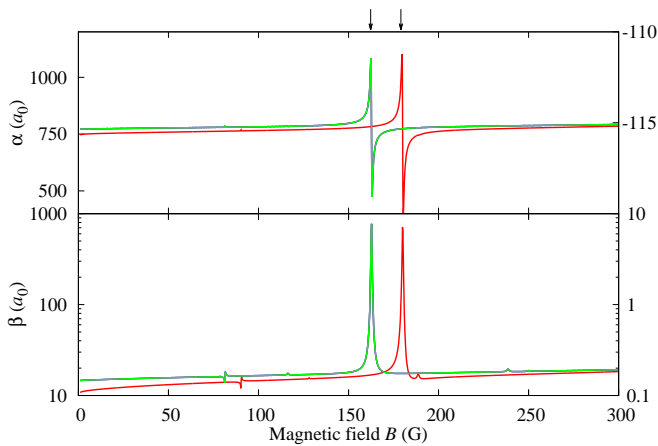}
\caption{[Color online.] The real and imaginary parts of the complex
scattering length at the Rb $|2,-1\rangle$ + Cs $|3,3\rangle$
threshold, calculated using the final fitted potentials: $s$-wave (red, dark gray, left-hand axis) and $p$-wave (green, light gray, right-hand axis). The observed resonance positions are shown as arrows at the top of the graph. The
three components of the $p$-wave scattering length are
indistinguishable on this scale.} \label{ea-scat}
\end{figure}

In addition to the resonances at the lowest threshold, Pilch {\em et
al.}\ \cite{Pilch:2009} observed two resonances at an excited threshold
with the Rb atoms in their $|2,-1\rangle$ state, at 162.3 and 179.1~G.
At this threshold inelastic scattering is possible, and trap loss can
occur through either 2-body or 3-body collisions. The scattering length
is complex, $a(B)=\alpha(B)-i\beta(B)$, and the inelastic collision
rate is proportional to $\beta(B)$. Fig.~\ref{ea-scat} shows the real
and imaginary parts of the $s$-wave and $p$-wave scattering lengths at
this threshold, calculated using the fitted potentials, and compares
them to the experimental resonance positions. It may be seen that the
two observed resonances are in good agreement with the calculation,
with the high-field resonance arising from $s$-wave scattering and the
low-field resonance from $p$-wave scattering.

\subsubsection{Unassigned resonance}

As noted above, there is one resonance observed in $s$-wave scattering,
at 273.45 G, that does not appear in coupled-channel calculations on
the fitted potential using a basis set with $L_{\rm max}=2$. However,
there are numerous additional resonances that appear when basis sets
including more partial waves are used. In particular, a calculation
including $L=4$ functions yields an additional resonance at 275.07 G
that arises from the $\left|-6\,(2,4)\,g\,(-2,4)\right\rangle$ state.
The exact position of this resonance is quite sensitive to variations
of the potential within its uncertainty and is plausibly
responsible for the otherwise unassigned resonance.

\subsubsection{High-field scattering}

\begin{figure}[tbp]
\vspace*{1ex}
\includegraphics[angle=-90,width=\columnwidth]{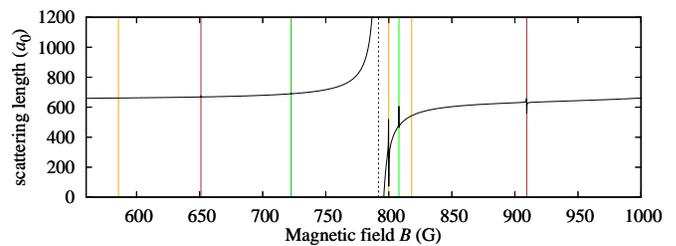}
\caption{[Color online.] RbCs scattering length at the
$|1,1\rangle + |3,3\rangle$ threshold at fields above 560 G, calculated
 at $E=160$ nK using the final fitted potential. Resonance positions are
marked by vertical lines, with the value of $M_F$ of the corresponding
bound state indicated using the same color scheme as in
Fig.~\ref{s-wave-low}.} \label{s-wave-high}
\end{figure}

The resonances listed in Table \ref{quality-of-fit}, at fields up to
553 G, include all those expected from $|{-6}(2,4)d\rangle$ states.
However, there are additional resonances that appear at higher field,
mostly due to $s$ and $d$ states of $|{-5}(2,3)\rangle$. Some of the
corresponding bound states appear in Fig.~\ref{s-wave-low}. Figure
\ref{s-wave-high} shows the $s$-wave scattering length at fields up to
1000 G; in particular, the comparatively wide resonance near 790 G
(with width $\Delta=4.2$~G) is due to the $|{-5}(2,3)s(2,2)\rangle$
state. This wide resonance may be useful for tuning interspecies
scattering properties, and for studying few-body properties such as
interspecies Efimov resonances \cite{Kraemer:2006}. In particular,
since there is a very broad Feshbach resonance for Cs in state
$|3,3\rangle$ with a pole at $787$~G \cite{Berninger:Efimov:2011},
Rb+Cs mixtures may make it possible to study Efimov physics near
overlapping Feshbach resonances.

\section{Outlook}
We have studied and modeled interspecies scattering in an ultracold
Rb-Cs gas mixture with the aim of finding an assignment for the
observed interspecies Feshbach resonances and in particular to
understand the spectrum of weakly bound RbCs molecules.

Our results are of great importance for the production of ultracold
samples of heteronuclear molecules and for the generation of dipolar
quantum gases made of RbCs molecules. With recent work on optical one-
and two-photon spectroscopy \cite{Debatin:2011} we are now poised to
perform stimulated ground-state transfer using the STIRAP technique. We
expect that a three-dimensional optical lattice will allow us to
maximize the molecule creation and state transfer efficiencies, as in
recent work on Cs$_2$ \cite{Danzl:ground:2010}. As detailed in
Ref.~\cite{Lercher:2011}, interspecies Feshbach tuning will be used to
bring a superfluid sample of Rb atoms into overlap with a
single-atom-per-site Mott insulator for Cs, in order to optimize the
Rb-Cs pair-creation efficiency. With sufficiently high efficiencies,
the creation of a dipolar quantum gas of RbCs molecules is within
reach.

\begin{acknowledgments}
The Innsbruck team acknowledges support by the Austrian Science Fund
(FWF) and the European Science Foundation (ESF) within the
EuroQUAM/QuDipMol project (FWF project number I124-N16) and support by
the FWF through the SFB FoQuS (FWF project number F4006-N16). The
Durham, JQI and Temple University teams acknowledge support from an
AFOSR MURI project on Ultracold Polar Molecules. The Durham and JQI
teams acknowledge support from the Engineering and Physical Sciences
Research Council. Work at Temple University was also supported by NSF
Grant PHY 1005453. The Durham and Hannover teams acknowledge support
from the QuDipMol project and the Hannover team also support by the
Deutsche Forschungsgemeinschaft through the cluster of excellence
QUEST.
\end{acknowledgments}

\bibliographystyle{apsrev}
%\bibliography{FeshbachAndWiggle,ultracold,all}

\begin{thebibliography}{67}
\expandafter\ifx\csname natexlab\endcsname\relax\def\natexlab#1{#1}\fi
\expandafter\ifx\csname bibnamefont\endcsname\relax
  \def\bibnamefont#1{#1}\fi
\expandafter\ifx\csname bibfnamefont\endcsname\relax
  \def\bibfnamefont#1{#1}\fi
\expandafter\ifx\csname citenamefont\endcsname\relax
  \def\citenamefont#1{#1}\fi
\expandafter\ifx\csname url\endcsname\relax
  \def\url#1{\texttt{#1}}\fi
\expandafter\ifx\csname urlprefix\endcsname\relax\def\urlprefix{URL }\fi
\providecommand{\bibinfo}[2]{#2}
\providecommand{\eprint}[2][]{\url{#2}}

\bibitem[{\citenamefont{Zwierlein et~al.}(2005)\citenamefont{Zwierlein,
  Abo-Shaeer, Schirotzek, Schunck, and Ketterle}}]{Zwierlein:2005}
\bibinfo{author}{\bibfnamefont{M.~W.} \bibnamefont{Zwierlein}},
  \bibinfo{author}{\bibfnamefont{J.~R.} \bibnamefont{Abo-Shaeer}},
  \bibinfo{author}{\bibfnamefont{A.}~\bibnamefont{Schirotzek}},
  \bibinfo{author}{\bibfnamefont{C.~H.} \bibnamefont{Schunck}},
  \bibnamefont{and} \bibinfo{author}{\bibfnamefont{W.}~\bibnamefont{Ketterle}},
  \bibinfo{journal}{Nature} \textbf{\bibinfo{volume}{435}},
  \bibinfo{pages}{1047} (\bibinfo{year}{2005}).

\bibitem[{\citenamefont{Greiner et~al.}(2002)\citenamefont{Greiner, Mandel,
  Esslinger, H\"{a}nsch, and Bloch}}]{Greiner:2002}
\bibinfo{author}{\bibfnamefont{M.}~\bibnamefont{Greiner}},
  \bibinfo{author}{\bibfnamefont{O.}~\bibnamefont{Mandel}},
  \bibinfo{author}{\bibfnamefont{T.}~\bibnamefont{Esslinger}},
  \bibinfo{author}{\bibfnamefont{T.~W.} \bibnamefont{H\"{a}nsch}},
  \bibnamefont{and} \bibinfo{author}{\bibfnamefont{I.}~\bibnamefont{Bloch}},
  \bibinfo{journal}{Nature} \textbf{\bibinfo{volume}{415}}, \bibinfo{pages}{39}
  (\bibinfo{year}{2002}).

\bibitem[{\citenamefont{Giorgini et~al.}(2008)\citenamefont{Giorgini,
  Pitaevskii, and Stringari}}]{Giorgini:2008}
\bibinfo{author}{\bibfnamefont{S.}~\bibnamefont{Giorgini}},
  \bibinfo{author}{\bibfnamefont{L.~P.} \bibnamefont{Pitaevskii}},
  \bibnamefont{and}
  \bibinfo{author}{\bibfnamefont{S.}~\bibnamefont{Stringari}},
  \bibinfo{journal}{Rev. Mod. Phys.} \textbf{\bibinfo{volume}{80}},
  \bibinfo{pages}{1215} (\bibinfo{year}{2008}).

\bibitem[{\citenamefont{Griesmaier et~al.}(2005)\citenamefont{Griesmaier,
  Werner, Hensler, Stuhler, and Pfau}}]{Griesmaier:2005}
\bibinfo{author}{\bibfnamefont{A.}~\bibnamefont{Griesmaier}},
  \bibinfo{author}{\bibfnamefont{J.}~\bibnamefont{Werner}},
  \bibinfo{author}{\bibfnamefont{S.}~\bibnamefont{Hensler}},
  \bibinfo{author}{\bibfnamefont{J.}~\bibnamefont{Stuhler}}, \bibnamefont{and}
  \bibinfo{author}{\bibfnamefont{T.}~\bibnamefont{Pfau}},
  \bibinfo{journal}{Phys. Rev. Lett.} \textbf{\bibinfo{volume}{94}},
  \bibinfo{pages}{160401} (\bibinfo{year}{2005}).

\bibitem[{\citenamefont{Griesmaier et~al.}(2006)\citenamefont{Griesmaier,
  Stuhler, and Pfau}}]{Griesmaier:2006}
\bibinfo{author}{\bibfnamefont{A.}~\bibnamefont{Griesmaier}},
  \bibinfo{author}{\bibfnamefont{J.}~\bibnamefont{Stuhler}}, \bibnamefont{and}
  \bibinfo{author}{\bibfnamefont{T.}~\bibnamefont{Pfau}},
  \bibinfo{journal}{Appl. Phys. B -- Lasers Opt.}
  \textbf{\bibinfo{volume}{82}}, \bibinfo{pages}{211} (\bibinfo{year}{2006}).

\bibitem[{\citenamefont{Ni et~al.}(2008)\citenamefont{Ni, Ospelkaus, {de
  Miranda}, Pe'er, Neyenhuis, Zirbel, Kotochigova, Julienne, Jin, and
  Ye}}]{Ni:KRb:2008}
\bibinfo{author}{\bibfnamefont{K.-K.} \bibnamefont{Ni}},
  \bibinfo{author}{\bibfnamefont{S.}~\bibnamefont{Ospelkaus}},
  \bibinfo{author}{\bibfnamefont{M.~H.~G.} \bibnamefont{{de Miranda}}},
  \bibinfo{author}{\bibfnamefont{A.}~\bibnamefont{Pe'er}},
  \bibinfo{author}{\bibfnamefont{B.}~\bibnamefont{Neyenhuis}},
  \bibinfo{author}{\bibfnamefont{J.~J.} \bibnamefont{Zirbel}},
  \bibinfo{author}{\bibfnamefont{S.}~\bibnamefont{Kotochigova}},
  \bibinfo{author}{\bibfnamefont{P.~S.} \bibnamefont{Julienne}},
  \bibinfo{author}{\bibfnamefont{D.~S.} \bibnamefont{Jin}}, \bibnamefont{and}
  \bibinfo{author}{\bibfnamefont{J.}~\bibnamefont{Ye}},
  \bibinfo{journal}{Science} \textbf{\bibinfo{volume}{322}},
  \bibinfo{pages}{231} (\bibinfo{year}{2008}).

\bibitem[{\citenamefont{G\'{o}ral et~al.}(2002)\citenamefont{G\'{o}ral, Santos,
  and Lewenstein}}]{Goral:2002}
\bibinfo{author}{\bibfnamefont{K.}~\bibnamefont{G\'{o}ral}},
  \bibinfo{author}{\bibfnamefont{L.}~\bibnamefont{Santos}}, \bibnamefont{and}
  \bibinfo{author}{\bibfnamefont{M.}~\bibnamefont{Lewenstein}},
  \bibinfo{journal}{Phys. Rev. Lett.} \textbf{\bibinfo{volume}{88}},
  \bibinfo{pages}{170406} (\bibinfo{year}{2002}).

\bibitem[{\citenamefont{B\"uchler et~al.}(2007)\citenamefont{B\"uchler, Demler,
  Lukin, Micheli, Prokof'ev, Pupillo, and Zoller}}]{Buechler:2007}
\bibinfo{author}{\bibfnamefont{H.~P.} \bibnamefont{B\"uchler}},
  \bibinfo{author}{\bibfnamefont{E.}~\bibnamefont{Demler}},
  \bibinfo{author}{\bibfnamefont{M.}~\bibnamefont{Lukin}},
  \bibinfo{author}{\bibfnamefont{A.}~\bibnamefont{Micheli}},
  \bibinfo{author}{\bibfnamefont{N.}~\bibnamefont{Prokof'ev}},
  \bibinfo{author}{\bibfnamefont{G.}~\bibnamefont{Pupillo}}, \bibnamefont{and}
  \bibinfo{author}{\bibfnamefont{P.}~\bibnamefont{Zoller}},
  \bibinfo{journal}{Phys. Rev. Lett.} \textbf{\bibinfo{volume}{98}},
  \bibinfo{pages}{060404} (\bibinfo{year}{2007}).

\bibitem[{\citenamefont{Micheli et~al.}(2007)\citenamefont{Micheli, Pupillo,
  B\"uchler, and Zoller}}]{Micheli:2007}
\bibinfo{author}{\bibfnamefont{A.}~\bibnamefont{Micheli}},
  \bibinfo{author}{\bibfnamefont{G.}~\bibnamefont{Pupillo}},
  \bibinfo{author}{\bibfnamefont{H.~P.} \bibnamefont{B\"uchler}},
  \bibnamefont{and} \bibinfo{author}{\bibfnamefont{P.}~\bibnamefont{Zoller}},
  \bibinfo{journal}{Phys. Rev. A} \textbf{\bibinfo{volume}{76}},
  \bibinfo{pages}{043604} (\bibinfo{year}{2007}).

\bibitem[{\citenamefont{Pupillo et~al.}(2008)\citenamefont{Pupillo, Griessner,
  Micheli, Ortner, Wang, and Zoller}}]{Pupillo:2008}
\bibinfo{author}{\bibfnamefont{G.}~\bibnamefont{Pupillo}},
  \bibinfo{author}{\bibfnamefont{A.}~\bibnamefont{Griessner}},
  \bibinfo{author}{\bibfnamefont{A.}~\bibnamefont{Micheli}},
  \bibinfo{author}{\bibfnamefont{M.}~\bibnamefont{Ortner}},
  \bibinfo{author}{\bibfnamefont{D.-W.} \bibnamefont{Wang}}, \bibnamefont{and}
  \bibinfo{author}{\bibfnamefont{P.}~\bibnamefont{Zoller}},
  \bibinfo{journal}{Phys. Rev. Lett.} \textbf{\bibinfo{volume}{100}},
  \bibinfo{pages}{050402} (\bibinfo{year}{2008}).

\bibitem[{\citenamefont{Wall and Carr}(2009)}]{Wall:2009}
\bibinfo{author}{\bibfnamefont{M.~L.} \bibnamefont{Wall}} \bibnamefont{and}
  \bibinfo{author}{\bibfnamefont{L.~D.} \bibnamefont{Carr}},
  \bibinfo{journal}{New J. Phys.} \textbf{\bibinfo{volume}{11}},
  \bibinfo{pages}{055027} (\bibinfo{year}{2009}).

\bibitem[{\citenamefont{\.Zuchowski and Hutson}(2010)}]{Zuchowski:trimers:2010}
\bibinfo{author}{\bibfnamefont{P.~S.} \bibnamefont{\.Zuchowski}}
  \bibnamefont{and} \bibinfo{author}{\bibfnamefont{J.~M.}
  \bibnamefont{Hutson}}, \bibinfo{journal}{Phys. Rev. A}
  \textbf{\bibinfo{volume}{81}}, \bibinfo{pages}{060703(R)}
  (\bibinfo{year}{2010}).

\bibitem[{\citenamefont{Bethlem and Meijer}(2003)}]{Bethlem:IRPC:2003}
\bibinfo{author}{\bibfnamefont{H.~L.} \bibnamefont{Bethlem}} \bibnamefont{and}
  \bibinfo{author}{\bibfnamefont{G.}~\bibnamefont{Meijer}},
  \bibinfo{journal}{Int. Rev. Phys. Chem.} \textbf{\bibinfo{volume}{22}},
  \bibinfo{pages}{73} (\bibinfo{year}{2003}).

\bibitem[{\citenamefont{Sage et~al.}(2005)\citenamefont{Sage, Sainis, Bergeman,
  and DeMille}}]{Sage:2005}
\bibinfo{author}{\bibfnamefont{J.~M.} \bibnamefont{Sage}},
  \bibinfo{author}{\bibfnamefont{S.}~\bibnamefont{Sainis}},
  \bibinfo{author}{\bibfnamefont{T.}~\bibnamefont{Bergeman}}, \bibnamefont{and}
  \bibinfo{author}{\bibfnamefont{D.}~\bibnamefont{DeMille}},
  \bibinfo{journal}{Phys. Rev. Lett.} \textbf{\bibinfo{volume}{94}},
  \bibinfo{pages}{203001} (\bibinfo{year}{2005}).

\bibitem[{\citenamefont{Deiglmayr et~al.}(2008)\citenamefont{Deiglmayr,
  Grochola, Repp, M\"ortlbauer, Gl\"uck, Lange, Dulieu, Wester, and
  Weidem\"uller}}]{Deiglmayr:2008}
\bibinfo{author}{\bibfnamefont{J.}~\bibnamefont{Deiglmayr}},
  \bibinfo{author}{\bibfnamefont{A.}~\bibnamefont{Grochola}},
  \bibinfo{author}{\bibfnamefont{M.}~\bibnamefont{Repp}},
  \bibinfo{author}{\bibfnamefont{K.}~\bibnamefont{M\"ortlbauer}},
  \bibinfo{author}{\bibfnamefont{C.}~\bibnamefont{Gl\"uck}},
  \bibinfo{author}{\bibfnamefont{J.}~\bibnamefont{Lange}},
  \bibinfo{author}{\bibfnamefont{O.}~\bibnamefont{Dulieu}},
  \bibinfo{author}{\bibfnamefont{R.}~\bibnamefont{Wester}}, \bibnamefont{and}
  \bibinfo{author}{\bibfnamefont{M.}~\bibnamefont{Weidem\"uller}},
  \bibinfo{journal}{Phys. Rev. Lett.} \textbf{\bibinfo{volume}{101}},
  \bibinfo{pages}{133004} (\bibinfo{year}{2008}).

\bibitem[{\citenamefont{Aikawa et~al.}(2010)\citenamefont{Aikawa, Akamatsu,
  Hayashi, Oasa, Kobayashi, Naidon, Kishimoto, Ueda, and
  Inouye}}]{Aikawa2010cto}
\bibinfo{author}{\bibfnamefont{K.}~\bibnamefont{Aikawa}},
  \bibinfo{author}{\bibfnamefont{D.}~\bibnamefont{Akamatsu}},
  \bibinfo{author}{\bibfnamefont{M.}~\bibnamefont{Hayashi}},
  \bibinfo{author}{\bibfnamefont{K.}~\bibnamefont{Oasa}},
  \bibinfo{author}{\bibfnamefont{J.}~\bibnamefont{Kobayashi}},
  \bibinfo{author}{\bibfnamefont{P.}~\bibnamefont{Naidon}},
  \bibinfo{author}{\bibfnamefont{T.}~\bibnamefont{Kishimoto}},
  \bibinfo{author}{\bibfnamefont{M.}~\bibnamefont{Ueda}}, \bibnamefont{and}
  \bibinfo{author}{\bibfnamefont{S.}~\bibnamefont{Inouye}},
  \bibinfo{journal}{Phys. Rev. Lett.} \textbf{\bibinfo{volume}{105}},
  \bibinfo{pages}{203001} (\bibinfo{year}{2010}).

\bibitem[{\citenamefont{Regal et~al.}(2003)\citenamefont{Regal, Ticknor, Bohn,
  and Jin}}]{Regal2003cum}
\bibinfo{author}{\bibfnamefont{C.~A.} \bibnamefont{Regal}},
  \bibinfo{author}{\bibfnamefont{C.}~\bibnamefont{Ticknor}},
  \bibinfo{author}{\bibfnamefont{J.~L.} \bibnamefont{Bohn}}, \bibnamefont{and}
  \bibinfo{author}{\bibfnamefont{D.~S.} \bibnamefont{Jin}},
  \bibinfo{journal}{Nature} \textbf{\bibinfo{volume}{424}}, \bibinfo{pages}{47}
  (\bibinfo{year}{2003}).

\bibitem[{\citenamefont{Herbig et~al.}(2003)\citenamefont{Herbig, Kraemer,
  Mark, Weber, Chin, N\"agerl, and Grimm}}]{Herbig2003poa}
\bibinfo{author}{\bibfnamefont{J.}~\bibnamefont{Herbig}},
  \bibinfo{author}{\bibfnamefont{T.}~\bibnamefont{Kraemer}},
  \bibinfo{author}{\bibfnamefont{M.}~\bibnamefont{Mark}},
  \bibinfo{author}{\bibfnamefont{T.}~\bibnamefont{Weber}},
  \bibinfo{author}{\bibfnamefont{C.}~\bibnamefont{Chin}},
  \bibinfo{author}{\bibfnamefont{H.-C.} \bibnamefont{N\"agerl}},
  \bibnamefont{and} \bibinfo{author}{\bibfnamefont{R.}~\bibnamefont{Grimm}},
  \bibinfo{journal}{Science} \textbf{\bibinfo{volume}{301}},
  \bibinfo{pages}{1510} (\bibinfo{year}{2003}).

\bibitem[{\citenamefont{Bergmann et~al.}(1998)\citenamefont{Bergmann, Theuer,
  and Shore}}]{Bergmann1998cpt}
\bibinfo{author}{\bibfnamefont{K.}~\bibnamefont{Bergmann}},
  \bibinfo{author}{\bibfnamefont{H.}~\bibnamefont{Theuer}}, \bibnamefont{and}
  \bibinfo{author}{\bibfnamefont{B.~W.} \bibnamefont{Shore}},
  \bibinfo{journal}{Rev. Mod. Phys.} \textbf{\bibinfo{volume}{70}},
  \bibinfo{pages}{1003} (\bibinfo{year}{1998}).

\bibitem[{\citenamefont{Winkler et~al.}(2007)\citenamefont{Winkler, Lang,
  Thalhammer, {van der Straten}, Grimm, and {Hecker
  Denschlag}}}]{Winkler2007cot}
\bibinfo{author}{\bibfnamefont{K.}~\bibnamefont{Winkler}},
  \bibinfo{author}{\bibfnamefont{F.}~\bibnamefont{Lang}},
  \bibinfo{author}{\bibfnamefont{G.}~\bibnamefont{Thalhammer}},
  \bibinfo{author}{\bibfnamefont{P.}~\bibnamefont{{van der Straten}}},
  \bibinfo{author}{\bibfnamefont{R.}~\bibnamefont{Grimm}}, \bibnamefont{and}
  \bibinfo{author}{\bibfnamefont{J.}~\bibnamefont{{Hecker Denschlag}}},
  \bibinfo{journal}{Phys. Rev. Lett.} \textbf{\bibinfo{volume}{98}},
  \bibinfo{eid}{043201} (\bibinfo{year}{2007}).

\bibitem[{\citenamefont{Danzl et~al.}(2008)\citenamefont{Danzl, Haller,
  Gustavsson, Mark, Hart, Bouloufa, Dulieu, Ritsch, and
  N\"agerl}}]{Danzl2008qgo}
\bibinfo{author}{\bibfnamefont{J.~G.} \bibnamefont{Danzl}},
  \bibinfo{author}{\bibfnamefont{E.}~\bibnamefont{Haller}},
  \bibinfo{author}{\bibfnamefont{M.}~\bibnamefont{Gustavsson}},
  \bibinfo{author}{\bibfnamefont{M.~J.} \bibnamefont{Mark}},
  \bibinfo{author}{\bibfnamefont{R.}~\bibnamefont{Hart}},
  \bibinfo{author}{\bibfnamefont{N.}~\bibnamefont{Bouloufa}},
  \bibinfo{author}{\bibfnamefont{O.}~\bibnamefont{Dulieu}},
  \bibinfo{author}{\bibfnamefont{H.}~\bibnamefont{Ritsch}}, \bibnamefont{and}
  \bibinfo{author}{\bibfnamefont{H.-C.} \bibnamefont{N\"agerl}},
  \bibinfo{journal}{Science} \textbf{\bibinfo{volume}{321}},
  \bibinfo{pages}{1062} (\bibinfo{year}{2008}).

\bibitem[{\citenamefont{Lang et~al.}(2008)\citenamefont{Lang, Winkler, Strauss,
  Grimm, and {Hecker Denschlag}}}]{Lang2008utm}
\bibinfo{author}{\bibfnamefont{F.}~\bibnamefont{Lang}},
  \bibinfo{author}{\bibfnamefont{K.}~\bibnamefont{Winkler}},
  \bibinfo{author}{\bibfnamefont{C.}~\bibnamefont{Strauss}},
  \bibinfo{author}{\bibfnamefont{R.}~\bibnamefont{Grimm}}, \bibnamefont{and}
  \bibinfo{author}{\bibfnamefont{J.}~\bibnamefont{{Hecker Denschlag}}},
  \bibinfo{journal}{Phys. Rev. Lett.} \textbf{\bibinfo{volume}{101}},
  \bibinfo{eid}{133005} (\bibinfo{year}{2008}).

\bibitem[{\citenamefont{Mark et~al.}(2009)\citenamefont{Mark, Danzl, Haller,
  Gustavsson, Bouloufa, Dulieu, Salami, Bergeman, Ritsch, Hart
  et~al.}}]{Mark2009drf}
\bibinfo{author}{\bibfnamefont{M.~J.} \bibnamefont{Mark}},
  \bibinfo{author}{\bibfnamefont{J.~G.} \bibnamefont{Danzl}},
  \bibinfo{author}{\bibfnamefont{E.}~\bibnamefont{Haller}},
  \bibinfo{author}{\bibfnamefont{M.}~\bibnamefont{Gustavsson}},
  \bibinfo{author}{\bibfnamefont{N.}~\bibnamefont{Bouloufa}},
  \bibinfo{author}{\bibfnamefont{O.}~\bibnamefont{Dulieu}},
  \bibinfo{author}{\bibfnamefont{H.}~\bibnamefont{Salami}},
  \bibinfo{author}{\bibfnamefont{T.}~\bibnamefont{Bergeman}},
  \bibinfo{author}{\bibfnamefont{H.}~\bibnamefont{Ritsch}},
  \bibinfo{author}{\bibfnamefont{R.}~\bibnamefont{Hart}}, \bibnamefont{et~al.},
  \bibinfo{journal}{Appl. Phys. B} \textbf{\bibinfo{volume}{95}},
  \bibinfo{pages}{219} (\bibinfo{year}{2009}).

\bibitem[{\citenamefont{Danzl et~al.}(2010)\citenamefont{Danzl, Mark, Haller,
  Gustavsson, Hart, Aldegunde, Hutson, and N\"agerl}}]{Danzl:ground:2010}
\bibinfo{author}{\bibfnamefont{J.~G.} \bibnamefont{Danzl}},
  \bibinfo{author}{\bibfnamefont{M.~J.} \bibnamefont{Mark}},
  \bibinfo{author}{\bibfnamefont{E.}~\bibnamefont{Haller}},
  \bibinfo{author}{\bibfnamefont{M.}~\bibnamefont{Gustavsson}},
  \bibinfo{author}{\bibfnamefont{R.}~\bibnamefont{Hart}},
  \bibinfo{author}{\bibfnamefont{J.}~\bibnamefont{Aldegunde}},
  \bibinfo{author}{\bibfnamefont{J.~M.} \bibnamefont{Hutson}},
  \bibnamefont{and} \bibinfo{author}{\bibfnamefont{H.-C.}
  \bibnamefont{N\"agerl}}, \bibinfo{journal}{Nature Phys.}
  \textbf{\bibinfo{volume}{6}}, \bibinfo{pages}{265} (\bibinfo{year}{2010}).

\bibitem[{\citenamefont{Lercher et~al.}(2011)\citenamefont{Lercher, Takekoshi,
  Debatin, Schuster, Rameshan, Ferlaino, Grimm, and N\"agerl}}]{Lercher:2011}
\bibinfo{author}{\bibfnamefont{A.~D.} \bibnamefont{Lercher}},
  \bibinfo{author}{\bibfnamefont{T.}~\bibnamefont{Takekoshi}},
  \bibinfo{author}{\bibfnamefont{M.}~\bibnamefont{Debatin}},
  \bibinfo{author}{\bibfnamefont{B.}~\bibnamefont{Schuster}},
  \bibinfo{author}{\bibfnamefont{R.}~\bibnamefont{Rameshan}},
  \bibinfo{author}{\bibfnamefont{F.}~\bibnamefont{Ferlaino}},
  \bibinfo{author}{\bibfnamefont{R.}~\bibnamefont{Grimm}}, \bibnamefont{and}
  \bibinfo{author}{\bibfnamefont{H.-C.} \bibnamefont{N\"agerl}},
  \bibinfo{journal}{Eur. Phys. J. D} \textbf{\bibinfo{volume}{65}},
  \bibinfo{pages}{3} (\bibinfo{year}{2011}).

\bibitem[{\citenamefont{Debatin et~al.}(2011)\citenamefont{Debatin, Takekoshi,
  Rameshan, Reichs\"ollner, Ferlaino, Grimm, Vexiau, Bouloufa, Dulieu, and
  N\"agerl}}]{Debatin:2011}
\bibinfo{author}{\bibfnamefont{M.}~\bibnamefont{Debatin}},
  \bibinfo{author}{\bibfnamefont{T.}~\bibnamefont{Takekoshi}},
  \bibinfo{author}{\bibfnamefont{R.}~\bibnamefont{Rameshan}},
  \bibinfo{author}{\bibfnamefont{L.}~\bibnamefont{Reichs\"ollner}},
  \bibinfo{author}{\bibfnamefont{F.}~\bibnamefont{Ferlaino}},
  \bibinfo{author}{\bibfnamefont{R.}~\bibnamefont{Grimm}},
  \bibinfo{author}{\bibfnamefont{R.}~\bibnamefont{Vexiau}},
  \bibinfo{author}{\bibfnamefont{N.}~\bibnamefont{Bouloufa}},
  \bibinfo{author}{\bibfnamefont{O.}~\bibnamefont{Dulieu}}, \bibnamefont{and}
  \bibinfo{author}{\bibfnamefont{H.-C.} \bibnamefont{N\"agerl}},
  \bibinfo{journal}{Physical Chemistry Chemical Physics}
  \textbf{\bibinfo{volume}{13}}, \bibinfo{pages}{18926} (\bibinfo{year}{2011}).

\bibitem[{\citenamefont{Pilch et~al.}(2009)\citenamefont{Pilch, Lange,
  Prantner, Kerner, Ferlaino, N\"agerl, and Grimm}}]{Pilch:2009}
\bibinfo{author}{\bibfnamefont{K.}~\bibnamefont{Pilch}},
  \bibinfo{author}{\bibfnamefont{A.~D.} \bibnamefont{Lange}},
  \bibinfo{author}{\bibfnamefont{A.}~\bibnamefont{Prantner}},
  \bibinfo{author}{\bibfnamefont{G.}~\bibnamefont{Kerner}},
  \bibinfo{author}{\bibfnamefont{F.}~\bibnamefont{Ferlaino}},
  \bibinfo{author}{\bibfnamefont{H.-C.} \bibnamefont{N\"agerl}},
  \bibnamefont{and} \bibinfo{author}{\bibfnamefont{R.}~\bibnamefont{Grimm}},
  \bibinfo{journal}{Phys. Rev. A} \textbf{\bibinfo{volume}{79}},
  \bibinfo{pages}{042718} (\bibinfo{year}{2009}).

\bibitem[{\citenamefont{Docenko et~al.}(2011)\citenamefont{Docenko, Tamanis,
  Ferber, Kn\"ockel, and Tiemann}}]{Docenko:RbCs:2011}
\bibinfo{author}{\bibfnamefont{O.}~\bibnamefont{Docenko}},
  \bibinfo{author}{\bibfnamefont{M.}~\bibnamefont{Tamanis}},
  \bibinfo{author}{\bibfnamefont{R.}~\bibnamefont{Ferber}},
  \bibinfo{author}{\bibfnamefont{H.}~\bibnamefont{Kn\"ockel}},
  \bibnamefont{and} \bibinfo{author}{\bibfnamefont{E.}~\bibnamefont{Tiemann}},
  \bibinfo{journal}{Phys. Rev. A} \textbf{\bibinfo{volume}{83}},
  \bibinfo{pages}{052519} (\bibinfo{year}{2011}).

\bibitem[{\citenamefont{Derevianko et~al.}(2001)\citenamefont{Derevianko, Babb,
  and Dalgarno}}]{Derevianko:2001}
\bibinfo{author}{\bibfnamefont{A.}~\bibnamefont{Derevianko}},
  \bibinfo{author}{\bibfnamefont{J.~F.} \bibnamefont{Babb}}, \bibnamefont{and}
  \bibinfo{author}{\bibfnamefont{A.}~\bibnamefont{Dalgarno}},
  \bibinfo{journal}{Phys. Rev. A} \textbf{\bibinfo{volume}{63}},
  \bibinfo{pages}{052704} (\bibinfo{year}{2001}).

\bibitem[{\citenamefont{Porsev and Derevianko}(2003)}]{Porsev:2003}
\bibinfo{author}{\bibfnamefont{S.~G.} \bibnamefont{Porsev}} \bibnamefont{and}
  \bibinfo{author}{\bibfnamefont{A.}~\bibnamefont{Derevianko}},
  \bibinfo{journal}{J. Chem. Phys.} \textbf{\bibinfo{volume}{119}},
  \bibinfo{pages}{844} (\bibinfo{year}{2003}).

\bibitem[{\citenamefont{Chin et~al.}(2010)\citenamefont{Chin, Grimm, Tiesinga,
  and Julienne}}]{Chin:RMP:2010}
\bibinfo{author}{\bibfnamefont{C.}~\bibnamefont{Chin}},
  \bibinfo{author}{\bibfnamefont{R.}~\bibnamefont{Grimm}},
  \bibinfo{author}{\bibfnamefont{E.}~\bibnamefont{Tiesinga}}, \bibnamefont{and}
  \bibinfo{author}{\bibfnamefont{P.~S.} \bibnamefont{Julienne}},
  \bibinfo{journal}{Rev. Mod. Phys.} \textbf{\bibinfo{volume}{82}},
  \bibinfo{pages}{1225} (\bibinfo{year}{2010}).

\bibitem[{\citenamefont{Thompson et~al.}(2005)\citenamefont{Thompson, Hodby,
  and Wieman}}]{Thompson2005ump}
\bibinfo{author}{\bibfnamefont{S.~T.} \bibnamefont{Thompson}},
  \bibinfo{author}{\bibfnamefont{E.}~\bibnamefont{Hodby}}, \bibnamefont{and}
  \bibinfo{author}{\bibfnamefont{C.~E.} \bibnamefont{Wieman}},
  \bibinfo{journal}{Phys. Rev. Lett.} \textbf{\bibinfo{volume}{95}},
  \bibinfo{eid}{190404} (\bibinfo{year}{2005}).

\bibitem[{\citenamefont{Weber et~al.}(2008)\citenamefont{Weber, Barontini,
  Catani, Thalhammer, Inguscio, and Minardi}}]{Weber2008aou}
\bibinfo{author}{\bibfnamefont{C.}~\bibnamefont{Weber}},
  \bibinfo{author}{\bibfnamefont{G.}~\bibnamefont{Barontini}},
  \bibinfo{author}{\bibfnamefont{J.}~\bibnamefont{Catani}},
  \bibinfo{author}{\bibfnamefont{G.}~\bibnamefont{Thalhammer}},
  \bibinfo{author}{\bibfnamefont{M.}~\bibnamefont{Inguscio}}, \bibnamefont{and}
  \bibinfo{author}{\bibfnamefont{F.}~\bibnamefont{Minardi}},
  \bibinfo{journal}{Phys. Rev. A} \textbf{\bibinfo{volume}{78}},
  \bibinfo{pages}{061601} (\bibinfo{year}{2008}).

\bibitem[{\citenamefont{Lange et~al.}(2009)\citenamefont{Lange, Pilch,
  Prantner, Ferlaino, Engeser, N\"agerl, Grimm, and Chin}}]{Lange:2009}
\bibinfo{author}{\bibfnamefont{A.~D.} \bibnamefont{Lange}},
  \bibinfo{author}{\bibfnamefont{K.}~\bibnamefont{Pilch}},
  \bibinfo{author}{\bibfnamefont{A.}~\bibnamefont{Prantner}},
  \bibinfo{author}{\bibfnamefont{F.}~\bibnamefont{Ferlaino}},
  \bibinfo{author}{\bibfnamefont{B.}~\bibnamefont{Engeser}},
  \bibinfo{author}{\bibfnamefont{H.~C.} \bibnamefont{N\"agerl}},
  \bibinfo{author}{\bibfnamefont{R.}~\bibnamefont{Grimm}}, \bibnamefont{and}
  \bibinfo{author}{\bibfnamefont{C.}~\bibnamefont{Chin}},
  \bibinfo{journal}{Phys. Rev. A} \textbf{\bibinfo{volume}{79}},
  \bibinfo{pages}{013622} (\bibinfo{year}{2009}).

\bibitem[{\citenamefont{Beaufils et~al.}(2010)\citenamefont{Beaufils,
  Crubellier, Zanon, Laburthe-Tolra, Mar\`{e}chal, Vernac, and
  Gorceix}}]{Beaufils2010rfa}
\bibinfo{author}{\bibfnamefont{Q.}~\bibnamefont{Beaufils}},
  \bibinfo{author}{\bibfnamefont{A.}~\bibnamefont{Crubellier}},
  \bibinfo{author}{\bibfnamefont{T.}~\bibnamefont{Zanon}},
  \bibinfo{author}{\bibfnamefont{B.}~\bibnamefont{Laburthe-Tolra}},
  \bibinfo{author}{\bibfnamefont{E.}~\bibnamefont{Mar\`{e}chal}},
  \bibinfo{author}{\bibfnamefont{L.}~\bibnamefont{Vernac}}, \bibnamefont{and}
  \bibinfo{author}{\bibfnamefont{O.}~\bibnamefont{Gorceix}},
  \bibinfo{journal}{Eur. Phys. J. D} \textbf{\bibinfo{volume}{56}},
  \bibinfo{pages}{99} (\bibinfo{year}{2010}).

\bibitem[{\citenamefont{Cho et~al.}(2011)\citenamefont{Cho, McCarron, Jenkin,
  Koeppinger, and Cornish}}]{Cho2011}
\bibinfo{author}{\bibfnamefont{H.}~\bibnamefont{Cho}},
  \bibinfo{author}{\bibfnamefont{D.}~\bibnamefont{McCarron}},
  \bibinfo{author}{\bibfnamefont{D.}~\bibnamefont{Jenkin}},
  \bibinfo{author}{\bibfnamefont{M.}~\bibnamefont{Koeppinger}},
  \bibnamefont{and} \bibinfo{author}{\bibfnamefont{S.}~\bibnamefont{Cornish}},
  \bibinfo{journal}{Eur. Phys. J. D} \textbf{\bibinfo{volume}{65}},
  \bibinfo{pages}{125} (\bibinfo{year}{2011}).

\bibitem[{\citenamefont{McCarron et~al.}(2011)\citenamefont{McCarron, Cho,
  Jenkin, Koeppinger, and Cornish}}]{McCarron2011}
\bibinfo{author}{\bibfnamefont{D.}~\bibnamefont{McCarron}},
  \bibinfo{author}{\bibfnamefont{H.}~\bibnamefont{Cho}},
  \bibinfo{author}{\bibfnamefont{D.}~\bibnamefont{Jenkin}},
  \bibinfo{author}{\bibfnamefont{M.}~\bibnamefont{Koeppinger}},
  \bibnamefont{and} \bibinfo{author}{\bibfnamefont{S.}~\bibnamefont{Cornish}},
  \bibinfo{journal}{Phys. Rev. A} \textbf{\bibinfo{volume}{84}},
  \bibinfo{pages}{011603} (\bibinfo{year}{2011}).

\bibitem[{\citenamefont{Takekoshi et~al.}(1998)\citenamefont{Takekoshi,
  Patterson, and Knize}}]{Takekoshi:1998}
\bibinfo{author}{\bibfnamefont{T.}~\bibnamefont{Takekoshi}},
  \bibinfo{author}{\bibfnamefont{B.~M.} \bibnamefont{Patterson}},
  \bibnamefont{and} \bibinfo{author}{\bibfnamefont{R.~J.} \bibnamefont{Knize}},
  \bibinfo{journal}{Phys. Rev. Lett.} \textbf{\bibinfo{volume}{81}},
  \bibinfo{pages}{5105} (\bibinfo{year}{1998}).

\bibitem[{\citenamefont{Xu et~al.}(2003)\citenamefont{Xu, Mukaiyama,
  Abo-Shaeer, Chin, Miller, and Ketterle}}]{Xu:2003}
\bibinfo{author}{\bibfnamefont{K.}~\bibnamefont{Xu}},
  \bibinfo{author}{\bibfnamefont{T.}~\bibnamefont{Mukaiyama}},
  \bibinfo{author}{\bibfnamefont{J.~R.} \bibnamefont{Abo-Shaeer}},
  \bibinfo{author}{\bibfnamefont{J.~K.} \bibnamefont{Chin}},
  \bibinfo{author}{\bibfnamefont{D.~E.} \bibnamefont{Miller}},
  \bibnamefont{and} \bibinfo{author}{\bibfnamefont{W.}~\bibnamefont{Ketterle}},
  \bibinfo{journal}{Phys. Rev. Lett.} \textbf{\bibinfo{volume}{91}},
  \bibinfo{pages}{210402} (\bibinfo{year}{2003}).

\bibitem[{\citenamefont{Thalhammer et~al.}(2006)\citenamefont{Thalhammer,
  Winkler, Lang, Schmid, Grimm, and Hecker~Denschlag}}]{Thalhammer:2006}
\bibinfo{author}{\bibfnamefont{G.}~\bibnamefont{Thalhammer}},
  \bibinfo{author}{\bibfnamefont{K.}~\bibnamefont{Winkler}},
  \bibinfo{author}{\bibfnamefont{F.}~\bibnamefont{Lang}},
  \bibinfo{author}{\bibfnamefont{S.}~\bibnamefont{Schmid}},
  \bibinfo{author}{\bibfnamefont{R.}~\bibnamefont{Grimm}}, \bibnamefont{and}
  \bibinfo{author}{\bibfnamefont{J.}~\bibnamefont{Hecker~Denschlag}},
  \bibinfo{journal}{Phys. Rev. Lett.} \textbf{\bibinfo{volume}{96}},
  \bibinfo{pages}{050402} (\bibinfo{year}{2006}).

\bibitem[{\citenamefont{Mark et~al.}(2007)\citenamefont{Mark, Kraemer,
  Waldburger, Herbig, Chin, N\"{a}gerl, and Grimm}}]{Mark2007siw}
\bibinfo{author}{\bibfnamefont{M.}~\bibnamefont{Mark}},
  \bibinfo{author}{\bibfnamefont{T.}~\bibnamefont{Kraemer}},
  \bibinfo{author}{\bibfnamefont{P.}~\bibnamefont{Waldburger}},
  \bibinfo{author}{\bibfnamefont{J.}~\bibnamefont{Herbig}},
  \bibinfo{author}{\bibfnamefont{C.}~\bibnamefont{Chin}},
  \bibinfo{author}{\bibfnamefont{H.-C.} \bibnamefont{N\"{a}gerl}},
  \bibnamefont{and} \bibinfo{author}{\bibfnamefont{R.}~\bibnamefont{Grimm}},
  \bibinfo{journal}{Phys. Rev. Lett.} \textbf{\bibinfo{volume}{99}},
  \bibinfo{eid}{113201} (\bibinfo{year}{2007}).

\bibitem[{\citenamefont{Knoop et~al.}(2008)\citenamefont{Knoop, Mark, Ferlaino,
  Danzl, Kraemer, N\"{a}gerl, and Grimm}}]{Knoop2008mfm}
\bibinfo{author}{\bibfnamefont{S.}~\bibnamefont{Knoop}},
  \bibinfo{author}{\bibfnamefont{M.}~\bibnamefont{Mark}},
  \bibinfo{author}{\bibfnamefont{F.}~\bibnamefont{Ferlaino}},
  \bibinfo{author}{\bibfnamefont{J.~G.} \bibnamefont{Danzl}},
  \bibinfo{author}{\bibfnamefont{T.}~\bibnamefont{Kraemer}},
  \bibinfo{author}{\bibfnamefont{H.-C.} \bibnamefont{N\"{a}gerl}},
  \bibnamefont{and} \bibinfo{author}{\bibfnamefont{R.}~\bibnamefont{Grimm}},
  \bibinfo{journal}{Phys. Rev. Lett.} \textbf{\bibinfo{volume}{100}},
  \bibinfo{eid}{083002} (\bibinfo{year}{2008}).

\bibitem[{\citenamefont{Mark et~al.}(2005)\citenamefont{Mark, Kraemer, Herbig,
  Chin, N\"agerl, and Grimm}}]{Mark2005eco}
\bibinfo{author}{\bibfnamefont{M.}~\bibnamefont{Mark}},
  \bibinfo{author}{\bibfnamefont{T.}~\bibnamefont{Kraemer}},
  \bibinfo{author}{\bibfnamefont{J.}~\bibnamefont{Herbig}},
  \bibinfo{author}{\bibfnamefont{C.}~\bibnamefont{Chin}},
  \bibinfo{author}{\bibfnamefont{H.-C.} \bibnamefont{N\"agerl}},
  \bibnamefont{and} \bibinfo{author}{\bibfnamefont{R.}~\bibnamefont{Grimm}},
  \bibinfo{journal}{Europhys. Lett.} \textbf{\bibinfo{volume}{69}},
  \bibinfo{pages}{706} (\bibinfo{year}{2005}).

\bibitem[{\citenamefont{Mohr et~al.}(2007)\citenamefont{Mohr, Taylor, and
  Newell}}]{CODATA:2006}
\bibinfo{author}{\bibfnamefont{P.~J.} \bibnamefont{Mohr}},
  \bibinfo{author}{\bibfnamefont{B.~N.} \bibnamefont{Taylor}},
  \bibnamefont{and} \bibinfo{author}{\bibfnamefont{D.~B.}
  \bibnamefont{Newell}}, \emph{\bibinfo{title}{The 2006 {CODATA Recommended
  Values of the Fundamental Physical Constants, Web} version 5.1}},
  \bibinfo{howpublished}{National Institute of Standards and Technology,
  Gaithersburg, MD 20899} (\bibinfo{year}{2007}).

\bibitem[{\citenamefont{Bize et~al.}(1999)\citenamefont{Bize, Sortais, Santos,
  Mandache, Clairon, and Salomon}}]{Bize:1999}
\bibinfo{author}{\bibfnamefont{S.}~\bibnamefont{Bize}},
  \bibinfo{author}{\bibfnamefont{Y.}~\bibnamefont{Sortais}},
  \bibinfo{author}{\bibfnamefont{M.~S.} \bibnamefont{Santos}},
  \bibinfo{author}{\bibfnamefont{C.}~\bibnamefont{Mandache}},
  \bibinfo{author}{\bibfnamefont{A.}~\bibnamefont{Clairon}}, \bibnamefont{and}
  \bibinfo{author}{\bibfnamefont{C.}~\bibnamefont{Salomon}},
  \bibinfo{journal}{Europhys.\ Lett.} \textbf{\bibinfo{volume}{45}},
  \bibinfo{pages}{558} (\bibinfo{year}{1999}).

\bibitem[{\citenamefont{Pashov et~al.}(2007)\citenamefont{Pashov, Docenko,
  Tamanis, Ferber, Kn\"ockel, and Tiemann}}]{Pashov2007cot}
\bibinfo{author}{\bibfnamefont{A.}~\bibnamefont{Pashov}},
  \bibinfo{author}{\bibfnamefont{O.}~\bibnamefont{Docenko}},
  \bibinfo{author}{\bibfnamefont{M.}~\bibnamefont{Tamanis}},
  \bibinfo{author}{\bibfnamefont{R.}~\bibnamefont{Ferber}},
  \bibinfo{author}{\bibfnamefont{H.}~\bibnamefont{Kn\"ockel}},
  \bibnamefont{and} \bibinfo{author}{\bibfnamefont{E.}~\bibnamefont{Tiemann}},
  \bibinfo{journal}{Phys. Rev. A} \textbf{\bibinfo{volume}{76}},
  \bibinfo{pages}{022511} (\bibinfo{year}{2007}).

\bibitem[{\citenamefont{Tiesinga et~al.}(1998)\citenamefont{Tiesinga, Williams,
  and Julienne}}]{Tiesinga:na2:1998}
\bibinfo{author}{\bibfnamefont{E.}~\bibnamefont{Tiesinga}},
  \bibinfo{author}{\bibfnamefont{C.~J.} \bibnamefont{Williams}},
  \bibnamefont{and} \bibinfo{author}{\bibfnamefont{P.~S.}
  \bibnamefont{Julienne}}, \bibinfo{journal}{Phys. Rev. A}
  \textbf{\bibinfo{volume}{57}}, \bibinfo{pages}{4257} (\bibinfo{year}{1998}).

\bibitem[{\citenamefont{Hutson et~al.}(2008)\citenamefont{Hutson, Tiesinga, and
  Julienne}}]{Hutson:Cs2-note:2008}
\bibinfo{author}{\bibfnamefont{J.~M.} \bibnamefont{Hutson}},
  \bibinfo{author}{\bibfnamefont{E.}~\bibnamefont{Tiesinga}}, \bibnamefont{and}
  \bibinfo{author}{\bibfnamefont{P.~S.} \bibnamefont{Julienne}},
  \bibinfo{journal}{Phys. Rev. A} \textbf{\bibinfo{volume}{78}},
  \bibinfo{pages}{052703} (\bibinfo{year}{2008}), \bibinfo{note}{note that the
  matrix element of the dipolar spin-spin operator given in Eq.\ A2 of this
  paper omits a factor of $-\sqrt{30}$.}

\bibitem[{\citenamefont{Hutson and Green}(1994)}]{molscat:v14}
\bibinfo{author}{\bibfnamefont{J.~M.} \bibnamefont{Hutson}} \bibnamefont{and}
  \bibinfo{author}{\bibfnamefont{S.}~\bibnamefont{Green}},
  \emph{\bibinfo{title}{{MOLSCAT} computer program, version 14}},
  \bibinfo{howpublished}{distributed by Collaborative Computational Project
  No.\ 6 of the UK Engineering and Physical Sciences Research Council}
  (\bibinfo{year}{1994}).

\bibitem[{\citenamefont{Gonz\'{a}lez-Mart\'{\i}nez and
  Hutson}(2007)}]{Gonzalez-Martinez:2007}
\bibinfo{author}{\bibfnamefont{M.~L.} \bibnamefont{Gonz\'{a}lez-Mart\'{\i}nez}}
  \bibnamefont{and} \bibinfo{author}{\bibfnamefont{J.~M.}
  \bibnamefont{Hutson}}, \bibinfo{journal}{Phys. Rev. A}
  \textbf{\bibinfo{volume}{75}}, \bibinfo{pages}{022702}
  (\bibinfo{year}{2007}).

\bibitem[{\citenamefont{Manolopoulos}(1986)}]{Manolopoulos:1986}
\bibinfo{author}{\bibfnamefont{D.~E.} \bibnamefont{Manolopoulos}},
  \bibinfo{journal}{J. Chem. Phys.} \textbf{\bibinfo{volume}{85}},
  \bibinfo{pages}{6425} (\bibinfo{year}{1986}).

\bibitem[{\citenamefont{Alexander and Manolopoulos}(1987)}]{Alexander:1987}
\bibinfo{author}{\bibfnamefont{M.~H.} \bibnamefont{Alexander}}
  \bibnamefont{and} \bibinfo{author}{\bibfnamefont{D.~E.}
  \bibnamefont{Manolopoulos}}, \bibinfo{journal}{J. Chem. Phys.}
  \textbf{\bibinfo{volume}{86}}, \bibinfo{pages}{2044} (\bibinfo{year}{1987}).

\bibitem[{\citenamefont{Alexander}(1984)}]{Alexander:1984}
\bibinfo{author}{\bibfnamefont{M.~H.} \bibnamefont{Alexander}},
  \bibinfo{journal}{J. Chem. Phys.} \textbf{\bibinfo{volume}{81}},
  \bibinfo{pages}{4510} (\bibinfo{year}{1984}).

\bibitem[{\citenamefont{Johnson}(1973)}]{Johnson:1973}
\bibinfo{author}{\bibfnamefont{B.~R.} \bibnamefont{Johnson}},
  \bibinfo{journal}{J. Comput. Phys.} \textbf{\bibinfo{volume}{13}},
  \bibinfo{pages}{445} (\bibinfo{year}{1973}).

\bibitem[{\citenamefont{Hutson}(2007)}]{Hutson:res:2007}
\bibinfo{author}{\bibfnamefont{J.~M.} \bibnamefont{Hutson}},
  \bibinfo{journal}{New J. Phys.} \textbf{\bibinfo{volume}{9}},
  \bibinfo{pages}{152} (\bibinfo{year}{2007}).

\bibitem[{\citenamefont{Hutson}(1993)}]{Hutson:bound:1993}
\bibinfo{author}{\bibfnamefont{J.~M.} \bibnamefont{Hutson}},
  \emph{\bibinfo{title}{{BOUND} computer program, version 5}},
  \bibinfo{howpublished}{distributed by Collaborative Computational Project
  No.\ 6 of the UK Engineering and Physical Sciences Research Council}
  (\bibinfo{year}{1993}).

\bibitem[{\citenamefont{Hutson}(1994)}]{Hutson:CPC:1994}
\bibinfo{author}{\bibfnamefont{J.~M.} \bibnamefont{Hutson}},
  \bibinfo{journal}{Comput. Phys. Commun.} \textbf{\bibinfo{volume}{84}},
  \bibinfo{pages}{1} (\bibinfo{year}{1994}).

\bibitem[{\citenamefont{Smirnov and Chibisov}(1965)}]{Smirnov:1965}
\bibinfo{author}{\bibfnamefont{B.~M.} \bibnamefont{Smirnov}} \bibnamefont{and}
  \bibinfo{author}{\bibfnamefont{M.~I.} \bibnamefont{Chibisov}},
  \bibinfo{journal}{Sov. Phys. JETP} \textbf{\bibinfo{volume}{21}},
  \bibinfo{pages}{624} (\bibinfo{year}{1965}).

\bibitem[{\citenamefont{Stoof et~al.}(1988)\citenamefont{Stoof, Koelman, and
  Verhaar}}]{Stoof:1988}
\bibinfo{author}{\bibfnamefont{H.~T.~C.} \bibnamefont{Stoof}},
  \bibinfo{author}{\bibfnamefont{J.~M. V.~A.} \bibnamefont{Koelman}},
  \bibnamefont{and} \bibinfo{author}{\bibfnamefont{B.~J.}
  \bibnamefont{Verhaar}}, \bibinfo{journal}{Phys. Rev. B}
  \textbf{\bibinfo{volume}{38}}, \bibinfo{pages}{4688} (\bibinfo{year}{1988}).

\bibitem[{\citenamefont{Moerdijk et~al.}(1995)\citenamefont{Moerdijk, Verhaar,
  and Axelsson}}]{Moerdijk:1995}
\bibinfo{author}{\bibfnamefont{A.~J.} \bibnamefont{Moerdijk}},
  \bibinfo{author}{\bibfnamefont{B.~J.} \bibnamefont{Verhaar}},
  \bibnamefont{and} \bibinfo{author}{\bibfnamefont{A.}~\bibnamefont{Axelsson}},
  \bibinfo{journal}{Phys. Rev. A} \textbf{\bibinfo{volume}{51}},
  \bibinfo{pages}{4852} (\bibinfo{year}{1995}).

\bibitem[{\citenamefont{Mies et~al.}(1996)\citenamefont{Mies, Williams,
  Julienne, and Krauss}}]{Mies:1996}
\bibinfo{author}{\bibfnamefont{F.~H.} \bibnamefont{Mies}},
  \bibinfo{author}{\bibfnamefont{C.~J.} \bibnamefont{Williams}},
  \bibinfo{author}{\bibfnamefont{P.~S.} \bibnamefont{Julienne}},
  \bibnamefont{and} \bibinfo{author}{\bibfnamefont{M.}~\bibnamefont{Krauss}},
  \bibinfo{journal}{J. Res. Natl. Inst. Stand. Technol}
  \textbf{\bibinfo{volume}{101}}, \bibinfo{pages}{521} (\bibinfo{year}{1996}).

\bibitem[{\citenamefont{Kotochigova et~al.}(2000)\citenamefont{Kotochigova,
  Tiesinga, and Julienne}}]{Kotochigova:2001}
\bibinfo{author}{\bibfnamefont{S.}~\bibnamefont{Kotochigova}},
  \bibinfo{author}{\bibfnamefont{E.}~\bibnamefont{Tiesinga}}, \bibnamefont{and}
  \bibinfo{author}{\bibfnamefont{P.~S.} \bibnamefont{Julienne}},
  \bibinfo{journal}{Phys. Rev. A} \textbf{\bibinfo{volume}{63}},
  \bibinfo{pages}{012517} (\bibinfo{year}{2000}).

\bibitem[{\citenamefont{Kotochigova and Tiesinga}(2005)}]{Kotochigova:2005}
\bibinfo{author}{\bibfnamefont{S.}~\bibnamefont{Kotochigova}} \bibnamefont{and}
  \bibinfo{author}{\bibfnamefont{E.}~\bibnamefont{Tiesinga}},
  \bibinfo{journal}{J. Chem. Phys.} \textbf{\bibinfo{volume}{123}},
  \bibinfo{pages}{174304} (\bibinfo{year}{2005}).

\bibitem[{\citenamefont{Law and Hutson}(1997)}]{I-NoLLS}
\bibinfo{author}{\bibfnamefont{M.~M.} \bibnamefont{Law}} \bibnamefont{and}
  \bibinfo{author}{\bibfnamefont{J.~M.} \bibnamefont{Hutson}},
  \bibinfo{journal}{Comput. Phys. Commun.} \textbf{\bibinfo{volume}{102}},
  \bibinfo{pages}{252} (\bibinfo{year}{1997}).

\bibitem[{\citenamefont{Le~Roy}(1998)}]{LeRoy:1998}
\bibinfo{author}{\bibfnamefont{R.~J.} \bibnamefont{Le~Roy}},
  \bibinfo{journal}{J. Mol. Spectrosc.} \textbf{\bibinfo{volume}{191}},
  \bibinfo{pages}{223} (\bibinfo{year}{1998}).

\bibitem[{\citenamefont{Kraemer et~al.}(2006)\citenamefont{Kraemer, Mark,
  Waldburger, Danzl, Chin, Engeser, Lange, Pilch, Jaakkola, N\"{a}gerl
  et~al.}}]{Kraemer:2006}
\bibinfo{author}{\bibfnamefont{T.}~\bibnamefont{Kraemer}},
  \bibinfo{author}{\bibfnamefont{M.}~\bibnamefont{Mark}},
  \bibinfo{author}{\bibfnamefont{P.}~\bibnamefont{Waldburger}},
  \bibinfo{author}{\bibfnamefont{J.~G.} \bibnamefont{Danzl}},
  \bibinfo{author}{\bibfnamefont{C.}~\bibnamefont{Chin}},
  \bibinfo{author}{\bibfnamefont{B.}~\bibnamefont{Engeser}},
  \bibinfo{author}{\bibfnamefont{A.~D.} \bibnamefont{Lange}},
  \bibinfo{author}{\bibfnamefont{K.}~\bibnamefont{Pilch}},
  \bibinfo{author}{\bibfnamefont{A.}~\bibnamefont{Jaakkola}},
  \bibinfo{author}{\bibfnamefont{H.~C.} \bibnamefont{N\"{a}gerl}},
  \bibnamefont{et~al.}, \bibinfo{journal}{Nature}
  \textbf{\bibinfo{volume}{440}}, \bibinfo{pages}{315} (\bibinfo{year}{2006}).

\bibitem[{\citenamefont{Berninger et~al.}(2011)\citenamefont{Berninger,
  Zenesini, Huang, Harm, N\"agerl, Ferlaino, Grimm, Julienne, and
  Hutson}}]{Berninger:Efimov:2011}
\bibinfo{author}{\bibfnamefont{M.}~\bibnamefont{Berninger}},
  \bibinfo{author}{\bibfnamefont{A.}~\bibnamefont{Zenesini}},
  \bibinfo{author}{\bibfnamefont{B.}~\bibnamefont{Huang}},
  \bibinfo{author}{\bibfnamefont{W.}~\bibnamefont{Harm}},
  \bibinfo{author}{\bibfnamefont{H.-C.} \bibnamefont{N\"agerl}},
  \bibinfo{author}{\bibfnamefont{F.}~\bibnamefont{Ferlaino}},
  \bibinfo{author}{\bibfnamefont{R.}~\bibnamefont{Grimm}},
  \bibinfo{author}{\bibfnamefont{P.~S.} \bibnamefont{Julienne}},
  \bibnamefont{and} \bibinfo{author}{\bibfnamefont{J.~M.}
  \bibnamefont{Hutson}}, \bibinfo{journal}{Phys. Rev. Lett.}
  \textbf{\bibinfo{volume}{107}}, \bibinfo{pages}{120401}
  (\bibinfo{year}{2011}).

\end{thebibliography}

\end{document}